\documentclass[10pt,reqno,final]{article}
\usepackage[section]{placeins}
\usepackage{amsmath,amsfonts,amssymb,amsthm,version}
\usepackage{mathrsfs,fancybox,pifont}
\usepackage{graphicx}
\usepackage{url,hyperref}
\usepackage[notcite,notref]{showkeys}
\usepackage{color}
\usepackage{subfigure,multirow}
\usepackage{epstopdf}
\usepackage{cases}
\usepackage{mathtools}
\usepackage{algorithm,algorithmic}
\usepackage{authblk}
\usepackage{listings}
\usepackage{threeparttable}
\usepackage{caption}
\usepackage{subfigure}
\usepackage{longtable,booktabs}
\usepackage{indentfirst}
\usepackage{enumerate}
\usepackage{lipsum}

\allowdisplaybreaks

\numberwithin{equation}{section}
\numberwithin{figure}{section}
\numberwithin{table}{section}

\theoremstyle{plain}

\theoremstyle{definition}

\theoremstyle{remark}

\newcommand{\BibTeX}{{\rm B\kern-.05em{\sc i\kern-.025em b}\kern-.08em\ignorespaces
  T\kern-.1667em\lower.7ex\hbox{E}\kern-.125emX}}

\counterwithout{figure}{section}

\title{Music Influence Modeling Based on Directed Network Model \thanks{The work was supported in part by the Natural Science Foundation of China (621712432), in part by the Zhejiang Natural Science Foundation of China (LY19F020009,  LY19F010002), and in part by the 2022 Xinmiao Talent Program of Zhejiang Province. It was also sponsored by K.C.Wong Magna Fund in Ningbo University.

 (Corresponding author: Haiyong Xu)} }
\author[1]{Xuan Zhang \thanks{2496282868@qq.com}}
\author[1]{Tingdi Ren \thanks{tingdiren@gmail.com}}
\author[1]{Lihong Wang \thanks{wanglihong@nbu.edu.cn}}
\author[1]{Haiyong Xu \thanks{xuhaiyong@nbu.edu.cn}}
\affil[1]{School of Mathematics and Statistics, Ningbo University}

\date{}

\begin{document}

\maketitle

\begin{abstract}
 Studying the history of music may provide a glimpse into the development of human creativity as we examine the evolutionary and revolutionary trends in music and genres. First, a musical influence metric was created to construct a directed network of musical influence. Second, we examined the revolutions and development of musical genres, modeled the similarity, and explored similarities and influences within and between genres. Hierarchical cluster analysis and time series analysis of genres were used to explore the correlation between genres. Finally, Network Analysis, Semantic Analysis, and Random Forest Model are employed to find the revolutionaries.

The above work was applied to Country music to sort out and analyze its evolution. 
In studying the connection between music and the social environment,  time series analysis is used to determine the impact of social, political, or technological changes on music.


  \medskip
  \noindent{\bf Mathematics Subject Classification: } 
\end{abstract}

\section{Introduction}

Music has been a part of human society since ancient times and has become an important part of the cultural heritage \cite{maryprasith2000effects}. There are many factors that influence an artist when creating new music, including their natural creativity, current social, political events, the use of new instruments, and other personal experiences \cite{manturzewska1990biographical}. To understand the role of music in human experience, a music development assessment is needed to develop to understand and measure the impact of previously produced music on new music and musical artists. Some artists can list a dozen or more other artists who they say influenced their own musical work. Additionally, the influence can be measured by the degree of similarity between song characteristics (such as structure, rhythm, or lyrics.)

There are sometimes revolutionary shifts in music \cite{manuel1987marxism}, providing new sounds or rhythms, such as when new genres appear, or reinventing existing genres. This may be due to a series of small changes, the collaborative efforts by artists, a series of influential artists or changes within society. In addition, many music have similar melodies, and many artists have contributed to major shifts in a musical genre \cite{logan2003toward}. Sometimes these shifts are due to one artist influencing another. Sometimes it is a change that emerges in response to external events (such as major world events or technological advances). Thus, both the emergence of revolutionaries and similar music in the development of music are inseparable from the discussion of musical influence. It is meaningful to quantify and study the impact of music.

Recently, some works have also been done to measure the influence of music via constructing the similarity of theme  [5]-[9]. Compared to traditional retrieval, more current researches are based on tagging to measure music similarity, that is, to analyze the similarity of music by genre, artist, album, language, etc., through manual annotation.
Sanden et al. \cite{sanden2011empirical} analyzed the relationship between the acoustic characteristics of music and their tags. Hong \cite{hong2008tag} used user-annotated music tags to evaluate the similarity of songs, which is a tag-based classification of artists and music styles. Cyrillic Laurier \cite{laurier2009music} classified music into four categories: happy, sad, angry and tender, based on which techniques such as clustering are combined to achieve music-category mapping relationships.
More recently, with the development of computer technology, deep learning methods can better achieve music similarity assessment through feature extraction. Wolf \cite{tran2014feature} mainly used Restricted Boltzmann Machines(RBM) to feature preprocessing. Ezzaidi \cite{ezzaidi2006automatic} obtained music timbre features through Mel-Frequency Cepstral Coefficients (MFCCs), which has been widely studied and applied in this field due to its good results.

In the field of music style similarity detection, the commonly used machine learning are Hidden Markov Models \cite{shao2004unsupervised}, Linear Judgment Analysis \cite{lee2011music}, Support Vector Machines \cite{chen2010music}, Gaussian Mixture Models \cite{magno2007signal}, and K-means \cite{kim2007music}.

By analyzing the previous related researches, it is found that it basically determines the similarity and influence of music by constructing features or given one or more tags. However, these methods do not reflect the interactions between music and the revolutionaries in music development, which are often influential musicians \cite{nelson2004music}. There is relatively little work related to the study of revolutionaries in the development of music \cite{duguid2014revolutionaries}, so this is a gap in the field. How does one construct a model that is both a good measure of similar music in musical development and a good representation of the relationship between influences and influences while also representing the change agents in music?

Perhaps directed network models can solve these problems since directed networks have an essential role in some fields. E. A. Leicht et al.\cite{leicht2008community}successfully found communities or modules in directed networks. Kim \cite{kim2014community} used directed networks to analyze the social networks via community detection. Therefore, it might be an excellent choice to construct directed networks to the music genre.

To address the problems, we hope to better reflect the complex influence relationship in music development by building a directed network model to measure influence and similarity. The relevant literature shows that the current network model for music usually uses conventional topological networks to analyze the influence of music, only through some existing topological metrics. In contrast, the music influence network we constructed considers both similarity and influence metrics. First, we build a directed network by constructing music influence metrics, and then improve the original network by studying the nature of the subnetworks. Further, we define musical similarity to compare the similarity of musicians between genres and within genres. Finally, propose a method to find the revolutionaries from the network we built.

The comparison reveals that, compared with the conventional topological network, the influence metrics we constructed consider music's dissemination properties and has a better performance in the subnetwork analysis. In addition, the complex influence relationship between genres can be identified from the network we constructed. Finally, the proposed methods can further study the evolution and development of musical genres. Some indicators that can indicate the occurrence of revolution may be found, which is valuable for studying the underlying historical information.

The innovations and contributions of this paper are mainly reflected in the following aspects:
\begin{itemize}
\item This paper constructs the network to study music, and the model constructed can identify complex, influential relationships between genres. When studying the evolution of music genres, some complex social environment influences on music genres can be found by the proposed method.

\item Compared with the metrics constructed by machine learning,the metrics can be directly displayed on the network graph. Compared with the conventional topological network graph, the influence metrics constructed are based on the existing topological metrics and take into account the propagation nature of music, which performs better.

\item The proposed method can further investigate the process of evolution and development of musical genres. Some indicators that can indicate the occurrence of revolution may be found, which is valuable for studying the underlying historical information.

\end{itemize}

\section{The Proposed Method}

\begin{figure}[!h]
\centering
\includegraphics[width=11cm]{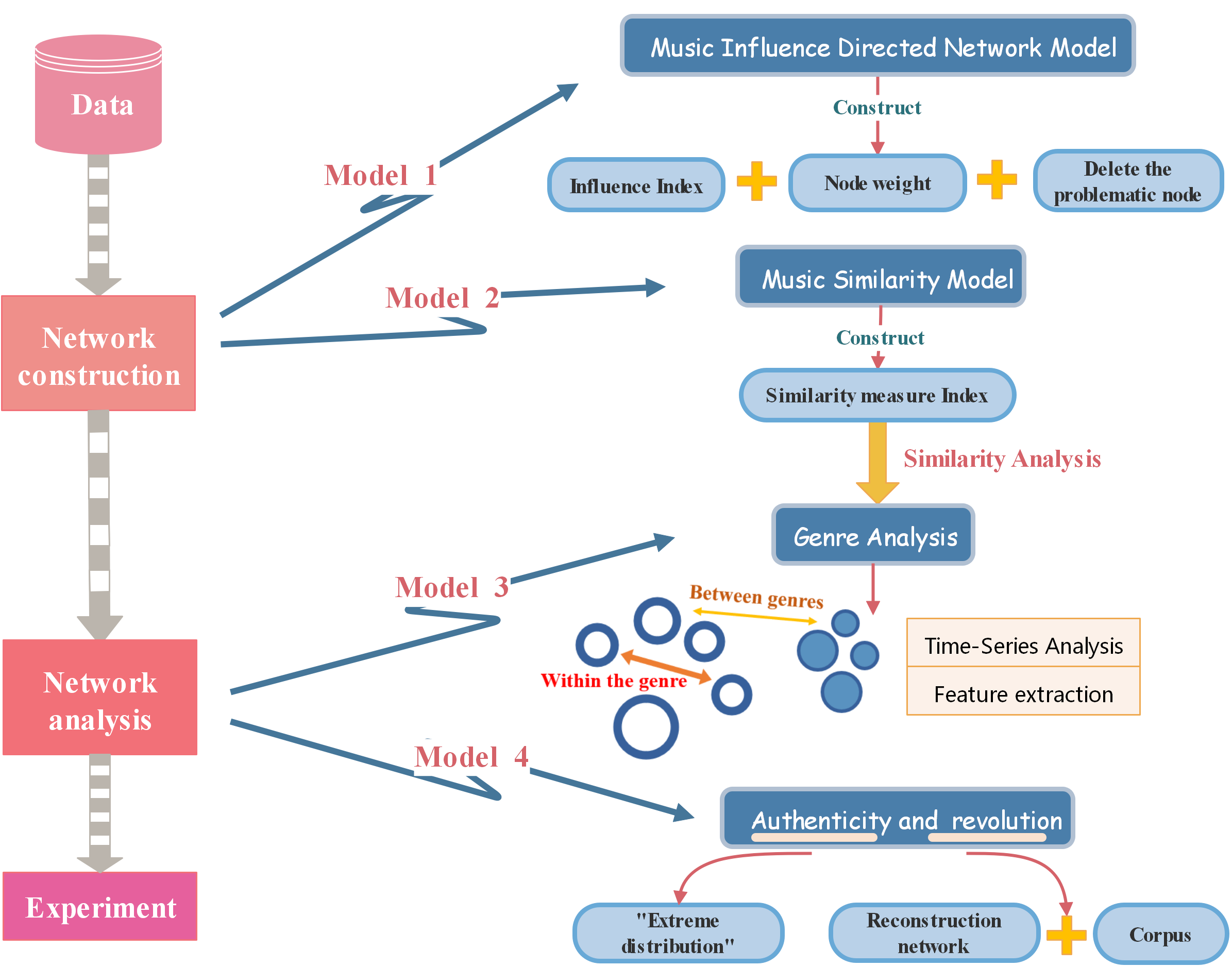}
\caption{The flowchart of the proposed method}
  \label{fig:11111}
\end{figure}

To better study the evolutionary trends of music and genres. The paper proposes four models as follows: Music Influence Directed Network Model(MIDN Model), Music Similarity Model(MS Model), Genre Analysis Model(GA Model), Influence Authenticity and Music Revolution Model(IAMR Model), respectively. 
The MIDN Model is designed to construct an influence metric to measure the size of a musician's influence and then initially build out the network. In MS Model, each song was considered a high-dimensional vector, and a method was proposed to measure high-dimensional vectors. 
In GA Model, the constructed influence and similarity metrics were drawn to compare influence and similarity within and between genres. 
In IAMR Model, "extreme distributions" are proposed to determine the authenticity of influence and find the revolutionaries in the development of music in the network.

The flow framework of the modeling is shown in Figure ~\ref{fig:11111}.
Experiments and discussions follow the modeling part, and the final part is the conclusion.

\subsection{Music Influence Directed Network Model}
\subsubsection{Analysis and modeling}
The data from \textbf{AllMusic.com} we collect will be used in building our network, which contains the influencers and followers of 5854 artists over the last 90 years. Based on these data, a direct network can be bulit by considering each musician as a node in a network. However, the first problem is whether the difference between the years of follower and influencer active can be used as the weight between the two nodes. In other words, whether the influence of an influencer is time-dependent. An influence metric is needed to answer this question.

Inspired by the epidemic, the following Figure ~\ref{fig:11}(a) will show a virus propagation event. It is well known that the most critical factor in virus propagation seems to be the speed of propagation, but the state of the virus on local and global should not be ignored. If there is an outbreak of some super-transmissible virus (e.g., Ebola) in an isolated village in the Amazon rainforest, it can say that the virus has the robust local transmission capability. However, the isolation factor causes the virus to have only a weak global transmission capability, that is, the global power status of the virus controls the spread of the virus, making its spread network converge and limited. Nevertheless, when the global power is immense, even if the virus has only weak local power and weak spread power, its dissemination network is bound to spread infinitely.

\begin{figure}[htbp]
\centering    
 \subfigure[Virus global-spread-local] 
{
	\begin{minipage}{7cm}
	\centering          
	\includegraphics[scale=0.35]{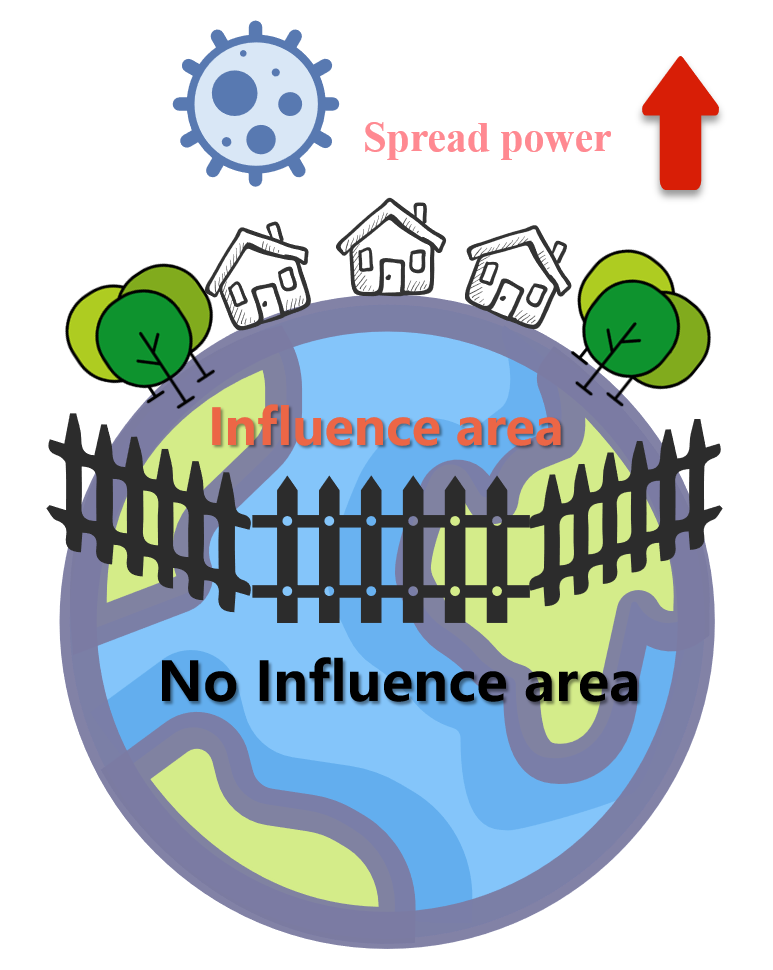}   
	\end{minipage}
}
	\subfigure[Index control] 
{
	\begin{minipage}{7cm}
	\centering      
	\includegraphics[scale=0.4]{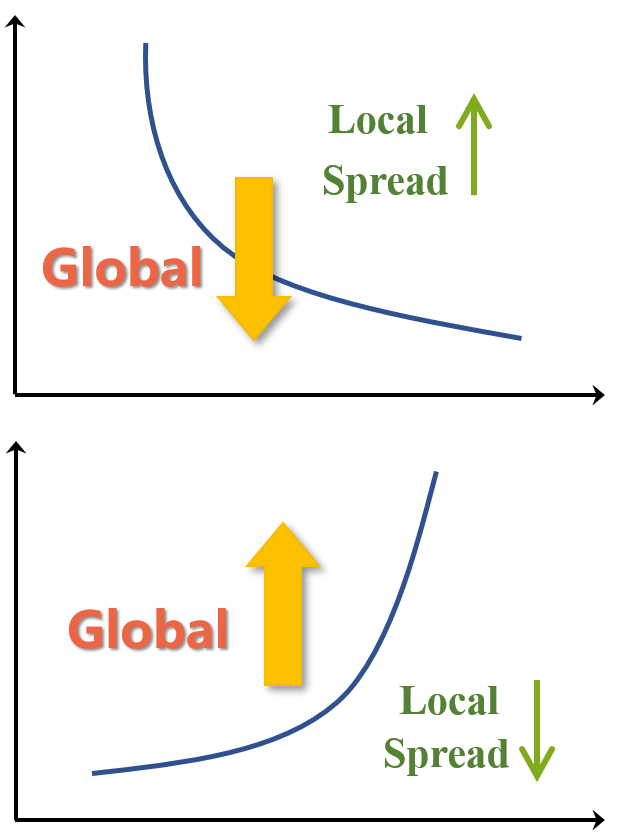}   
	\end{minipage}
}
 
\caption{name of the figure} 
\label{fig:11}  
\end{figure}

Nevertheless, for music networks, it is unrealistic to spread like a virus-like crazy, then it is feasible to control the product of local power and spread power composition with exponents, as shown in Figure ~\ref{fig:11}(b). Thus, it can be converted into a mathematical problem.

\textbf{Problem:}

Consider $x,y$, where $x,y>0$, and seek $f\left( x \right) $ to satisfy:
(1) \ $x\rightarrow 0,\ y\rightarrow \infty ,\ f\left( x \right) y\rightarrow 0$ \footnote{Infinity does not really tend to infinity, because the actual exponential values are always obtained in a finite range}

(2) \ $x\longrightarrow \infty ,\ y\longrightarrow 0,\ f\left( x \right) y\rightarrow \infty $  \footnote{The ideal case is $f\left( x \right) \cdot y\rightarrow x$, because the influence of the music network is always weak relative to the virus, but such a function is difficult to construct, we weaken the condition to turn to $f\left( x \right) \cdot y\rightarrow \infty $}


It is easy to construct the function to get $f\left( x \right) =e^x-1$, so the node influence score of the music network can be defined as: 

\begin{equation}\label{1}
NI_i\ =\ \left( e^{GC_i}-1 \right) LC_i\cdot SC_i  
\end{equation}  
where $LC_i,\ SC_i,\ GC_i$ represent Local Centrality, Spread Centrality, and Global Centrality, respectively. Below, we need specific measures of three centrality metrics.

For Local centrality, the ClusterRank metric established by PloS \cite{chen2013identifying} was chosen, which takes into account not only the number of neighbors and the neighbors' influences but also the clustering coefficient. $LC_i$, i.e. ClusterRank is express as:

\begin{equation}\label{2}
LC_i=f\left( c_i \right) \sum_{j\in N\left( i \right)}{\left( k_{j}^{\mathrm{out}}+1 \right)}
\end{equation}  
where the term $f\left( c_i \right) $ accounts for the effect of vertex $i$-th local clustering and $N\left( i \right) $ is the set of neighbors of vertex $i$ and the term $+1$ results from the contribution of $j$ itself.

For Spread centrality, we used the Semi-Local established by Chen et al. \cite{chen2012identifying}, which links an equilibrium of Local and Global centrality. They use the Susceptible-Infected-Recovered (SIR) model to evaluate the performance by using the spreading rate and the number of infected nodes.

\begin{equation}\label{3}
Q\left( u \right) =\sum_{i\in \Gamma \left( u \right)}{N\left( i \right)}
\end{equation}  

\begin{equation}\label{4}
SC_i=\sum_{i\in \Gamma \left( v \right)}{Q\left( u \right)}
\end{equation}  
where $\varGamma \left( u \right) $ is the set of the nearest neighbors of node $u$ and $N\left( i \right)$ is the number of the nearest and the next nearest neighbors of node $i$.

As for Global Centrality, considering as a musical directed network needs to focus on the output of the nodes,  OutCloesness is abled as the Global Centrality metric, which is the inversed sum of the distance from a node out to all other nodes in the graph.

\begin{equation}\label{5}
GC_i=\left( \frac{A_i}{N-1} \right) ^2\frac{1}{C_i}
\end{equation} 
where $C_i$ is the sum of distances from node $i$ to all reachable nodes.

With Eqs.(\ref{1}), (\ref{2}), (\ref{3}), (\ref{4}) and (\ref{5}), we successfully constructed the influence metric.

\begin{figure}[!h]
\centering
\includegraphics[width=8.8cm]{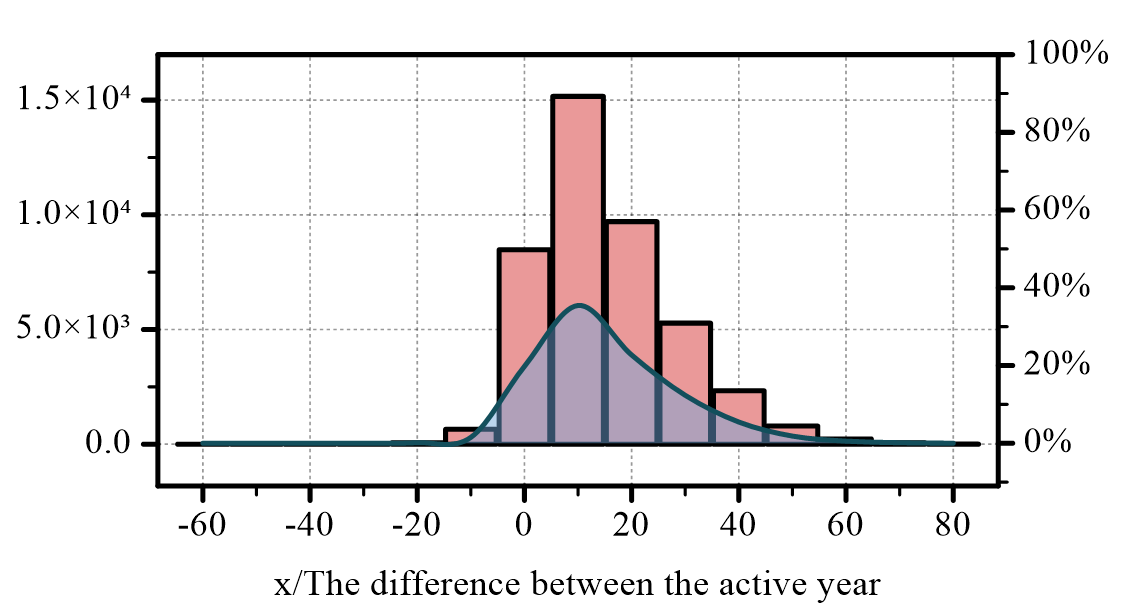}
\caption{Data distribution diagram of year difference}
  \label{fig:11-3}
\end{figure}

However, as a directed network, it needs to consider the weights between the nodes. Based on the collected data, we tried to construct the network with the year-difference as the weight, which is the difference between follower and influencer active years. Nevertheless, a considerable  fraction of non-positive data for the year-difference restricts ignorance. If the weight is zero, it means no connection between these two nodes; if the weight is negative, it is still plausible since it is common for the follower to be active earlier than the influencer in real life. Figure ~\ref{fig:11-3} shows the data distribution and proportions of year differences.

Therefore, Max-Min Normalization \cite{patel2011impact} is performed, with the sensitivity to extreme values, year-difference data satisfying  $\left\{ x|x\leqslant -30,x=80 \right\} $, which accounts for a relatively small, are ignored. In order to avoid zero values, the minimum value in normalization is defined as -30. The following transformation is available:
\begin{equation}\label{6}
Z=\frac{X-\left( -30 \right)}{X_{\max}-\left( -30 \right)},\ X\in \left[ -20,70 \right] 
\end{equation}

\begin{figure}[!h]
\centering
\includegraphics[width=5.6cm]{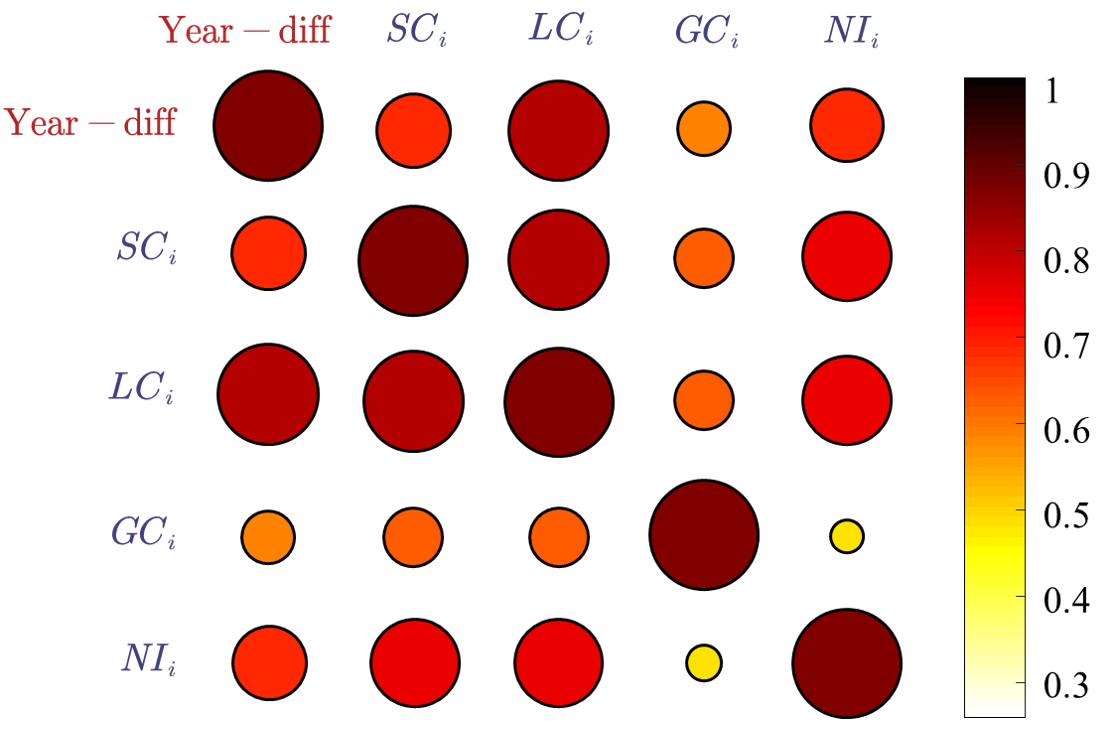}
\caption{Variable correlation coefficient of variables}
 \label{fig:11-4}
\end{figure}

Finally, the correlation coefficients between the year difference and each centrality measure are shown in Figure ~\ref{fig:11-4}.

Correlation coefficients indicate that the year-difference is significantly correlated with the centrality measure. Among them, the $LC_i$ metric strongly correlates with the year difference (0.9065). Therefore, it is reasonable to use the age difference as the weight of the network.
\subsubsection{Test of metrics measuring influence}

\begin{figure}[!h]
\centering
\includegraphics[width=15cm]{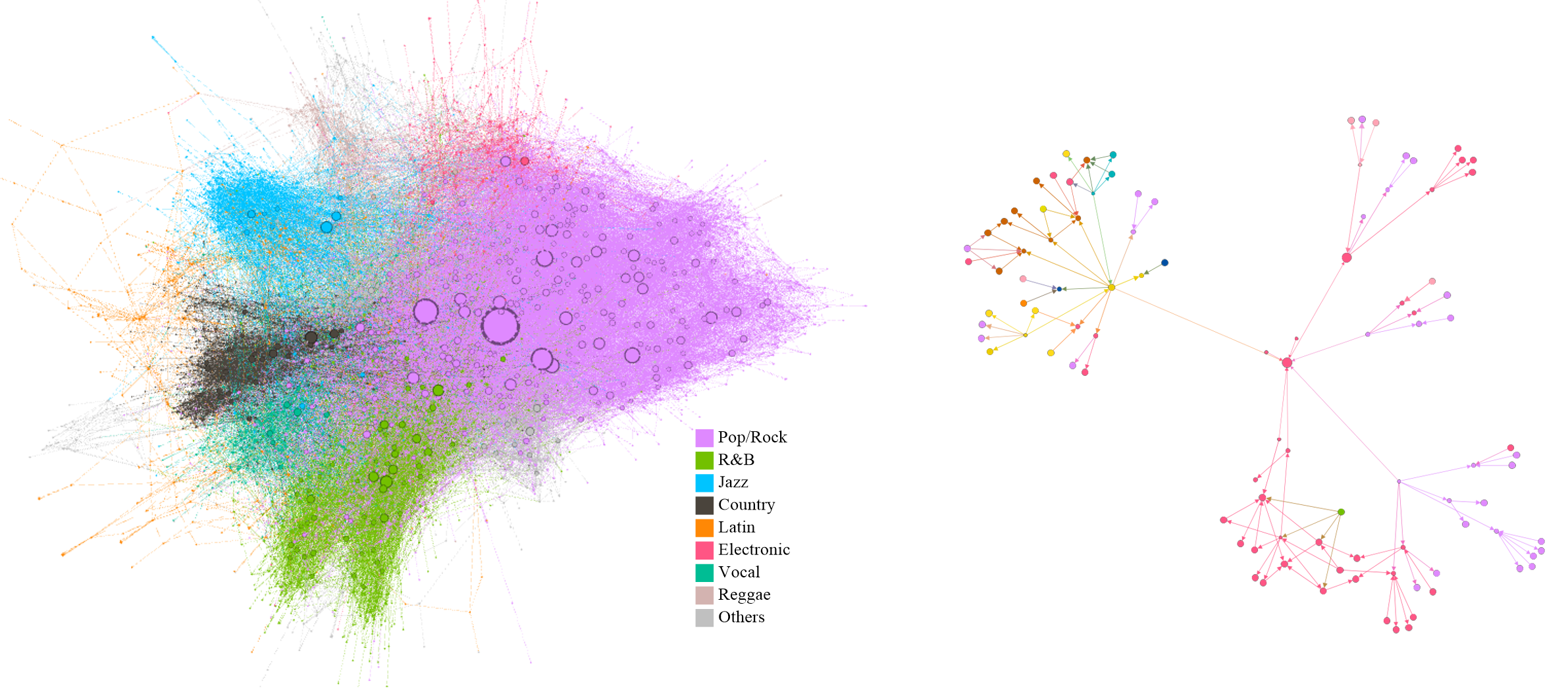}
\caption{Overview and partial maps of the directed music network}
 \label{fig:11-5}
\end{figure}

Based on the analysis in the previous section, the music influence measure $NI_i$ is regarded as the size attribute of the nodes, as well as the year-difference is used as the weight of the network. However, we found some sub-loops in the network, which may cause some influence on the metric. To avoid this problem, we \textbf{removed these sub-loops.} An overview diagram of the directed music network, as well as local diagrams, are shown in Figures ~\ref{fig:11-5} 
 
Top 5 influential musicians are filtered via $NI_i$. To check the reliability of the metric, it needs to combine the total number of people affected by the network and the authoritative evaluation of music status, as shown in Table 2.1.
\vspace{0.5cm}

\begin{table}[!h]
\caption{Top 5 influential musicians details}
\vspace{0.3cm}
\centering
\begin{threeparttable}          
\begin{tabular}{ccccc}

\hline
Rank\_NI & musician       & Merge    & \begin{tabular}[c]{@{}c@{}}Total number of \\ affected nodes\end{tabular} & Position in the music industry \tnote{1} \\ \hline
1        & The Beatles    & Pop/Rock & 4643                                                                      & \begin{tabular}[c]{@{}c@{}}The most influential \\ band of all time \cite{hasted2017you}                                                                    \end{tabular}                                                               \\ \hline
2        & Chuck Berry    & Pop/Rock & 4647                                                                      & \begin{tabular}[c]{@{}c@{}}One of the pioneers of\\  rock and roll music\end{tabular}                                                          \\ \hline
3        & Bob Dylan      & Pop/Rock & 4643                                                                      & \begin{tabular}[c]{@{}c@{}}One of the greatest \\ songwriters of all time\end{tabular}                                                         \\ \hline
4        & Hank Williams  & Country  & 4692                                                                      & \begin{tabular}[c]{@{}c@{}}One of the most significant \\  and influential American singers  \\ and songwriters of the 20th century\end{tabular} \\ \hline
5        & Little Richard & Pop/Rock & 4654                                                                      & \begin{tabular}[c]{@{}c@{}}Nicknamed "The Innovator, \\ The Originator, and The \\ Architect of Rock and Roll"\end{tabular}                    \\ \hline
\end{tabular}
\begin{tablenotes}    
        \footnotesize               
        \item[1] All reviews are from Wikipedia, except for the quoted content.          
      \end{tablenotes}            
    \end{threeparttable}       
\end{table}

\vspace{0.5cm}
Although the total number of influence nodes does not reflect the influence ranking well, the index is relatively reliable in terms of music industry status and public popularity.
\vspace{0.5cm}

\subsubsection{Index Analysis in Subnets}
The index is applied to a subnet composed of musicians from the Pop/Rock genre. We still searched for the Top 5 influential musicians in the Pop/Rock genre, whose information is shown in Table 2.2 below.

\vspace{0.5cm}
\begin{table}[!h]
\caption{Details of the top 5 musicians with influence in the Pop/Rock genre}
\vspace{0.3cm}
\begin{threeparttable}  
\begin{tabular}{ccccc}
\hline
Rank\_NI & Musician   & \begin{tabular}[c]{@{}c@{}}Number of first-order \\ affected nodes \tnote{1}   \end{tabular} & \begin{tabular}[c]{@{}c@{}}Number of second-order \\ affected nodes \tnote{2}    \end{tabular} & \begin{tabular}[c]{@{}c@{}}Total number of \\ affected nodes\end{tabular} \\ \hline
1      & The Beatles    & 615                                                                              & 2804                                                                              & 4643                                                                      \\ \hline
2     & Chuck Berry    & 159                                                                              & 1973                                                                              & 4647                                                                      \\ \hline
3     & Bob Dylan      & 389                                                                              & 2327                                                                              & 4643                                                                      \\ \hline
4     & Little Richard & 88                                                                               & 1910                                                                              & 4654                                                                      \\ \hline
5    & Elvis Presley  & 166                                                                              & 1977                                                                              & 4643                                                                      \\ \hline
\end{tabular}
\begin{tablenotes}    
        \footnotesize               
        \item[1] Number of first-order influence nodes: the number of nodes directly adjacent to the node
        \item[2] Number of second-order influence nodes: the number of nodes that are directly adjacent to the first-order influence node of the node.           
      \end{tablenotes}            
    \end{threeparttable}       
\end{table}

\begin{figure}[!h]
\centering
\includegraphics[width=11cm]{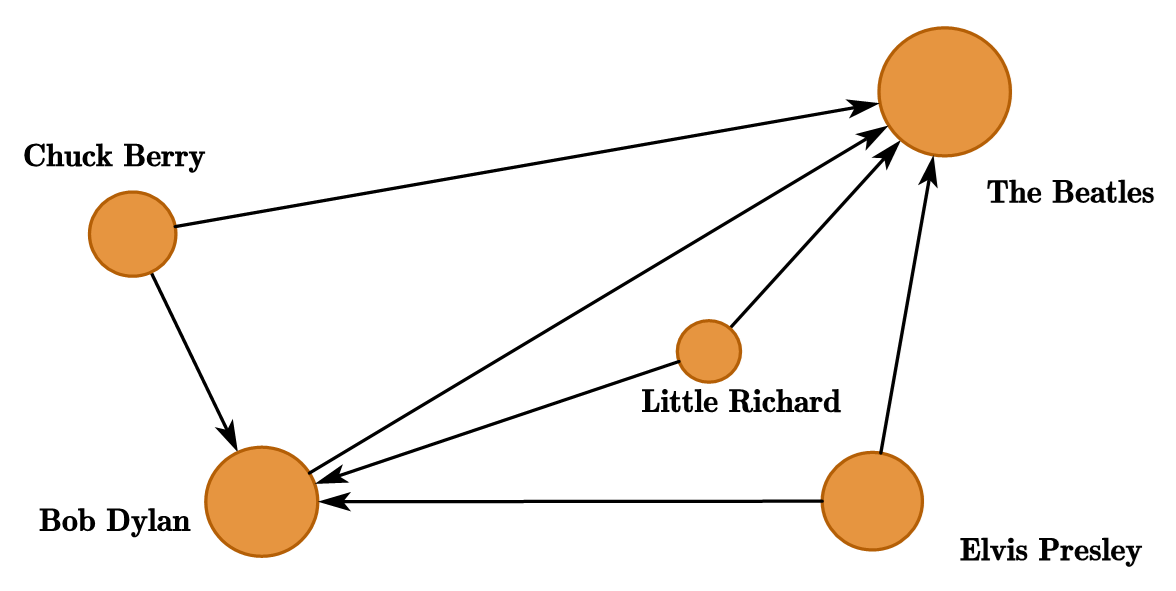}
\caption{Directed network diagram of Top 5 musicians}
  \label{fig:11-7}
\end{figure}

\newpage
There was little difference in the total number of affected nodes, but the number of first-order affected nodes differed much. Thus, these five musicians form a directed network graph, as shown in Figure ~\ref{fig:11-7}.

Obviously, the metric does not reflect the founders of the genre, i.e., the original nodes of the subnet, since Chuck Berry, Little Richard, and Elvis Presley all influenced The Beatles.

However, the metric is a good representation of the critical propagators of the genre, i.e., \textbf{ the key diffusion nodes of the subnetwork.} Both The Beatles and Bob Dylan are well known to the public, and by this criterion, The Beatles and Bob Dylan should be ranked higher than Chuck Berry. So, there is no doubt that The Beatles is Top 1. 
Although the total number of influence nodes of the three is almost the same, in terms of the number of first-order influence nodes, Chuck Berry is much smaller than others, suggesting that a significant portion of its total influence nodes comes from The Beatles and Bob Dylan. Nevertheless, Chuck Berry's emergence opened up the possibility of The Beatle's and Bob Dylan's emergence. The metrics proposed in this paper successfully capture the key opportunity's impact, which leads to the higher rank of Chuck Berry compared with Bob Dylan.

\vspace{0.3cm}
\subsection{Music Similarity Model}
\vspace{0.2cm}

The dataset used in this subsection of the model is from \textit{Spotify} API \cite{panda2021does}, which provides 16 variable entries, including musical characteristics such as Danceability, Tempo, Loudness, and Pitch, as well as Artist Name and Artist ID for 98,340 songs.

In the dataset, each song possesses multiple feature variables, and if these variables are considered as components of a vector, then all that needs to be done is to give metrics that can measure the similarity of the vector. Before constructing the indicators, these variables need to be processed and analyzed.

\vspace{0.2cm}
\subsubsection{Analysis and Processing of Variables}

First, descriptive statistics are needed for each variable to check for outliers, and then data outside the range of values present in the loudness variables are eliminated (the range of values required is $\left[ -60,0 \right] $).

The variables are divided into categorical and numeric variables, where Key and year can be considered numeric variables. For the variable Explicit, only 0.003\% of the songs have a value of 1, so the variable Explicit is discarded. In addition, it also need to consider the variable Mode, which is strongly correlated with Key in music theory and is derived from Key. Although a sample correlation test (due to the sheer size of the data) suggests that the two are only weakly correlated, this may be because the data are derived from estimates and some specific songs that are difficult to characterize. Therefore, the explicit variable is rejected.

\begin{figure}[!h]
\centering
\includegraphics[width=12.5cm]{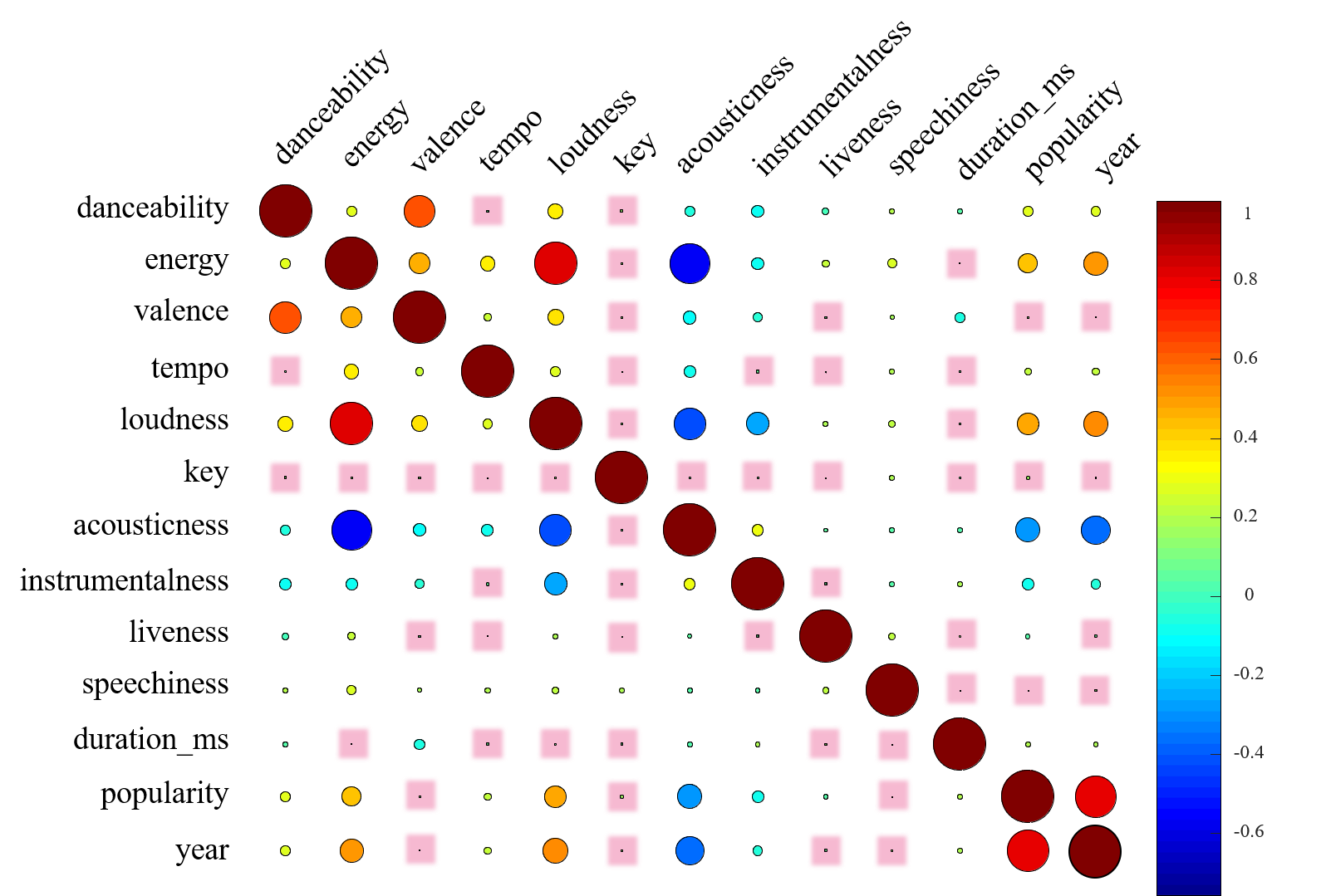}
\caption{Pearson correlation coefficient}
  \label{fig:22-1}
\end{figure}

The commonly used similarity metrics are Euclidean distance, Cosine similarity \cite{salton1988term}, but they do not necessarily perform well in the face of high dimensional vectors. A multicollinearity test was first performed to address the problem of too many characteristic variables. Due to a large amount of data in the design, we performed a random sample correlation analysis (1\% of the data sampled), and the results are shown in Figure ~\ref{fig:22-1} (pink background points indicate that the point is not significant at the 0.05 level).

Suppose the matrix formed by the original data is $X_{m\times n}$, and the $k$-dimensional matrix $Z_{m\times k}$ obtained after its dimensionality reduction, which can be tabulated by a mapping matrix, i.e., $Z_{m\times k}=X_{m\times k}W_{k\times n}$.  The purpose of dimensionality reduction is to make the distribution of the dimensioned data in the axis as dispersed as possible. And the dispersion of the data distribution can be measured by the variance,
so for $\boldsymbol{x}\in X,\ \boldsymbol{w}\in W,\ \boldsymbol{z}\in Z$, we can build a planning model:

\begin{equation}\label{7}
	\begin{split}
		\max_{\boldsymbol{w}} \frac{\boldsymbol{w}}{m}\cdot \mathrm{Cov}\left( \boldsymbol{x} \right) \boldsymbol{w}^T\\
		\mathrm{s}.\mathrm{t}. ||\boldsymbol{w}||_2=1
	\end{split}
\end{equation}

It can be solved using the Lagrangian function \cite{pistone2018lagrangian}, thus achieving dimensionality reduction and obtaining the component matrix (mapping matrix)$W$, as shown in Table 2.3.

\setlength{\tabcolsep}{7pt}
\begin{center}
\begin{longtable}{cccccccccc}
\caption{Component matrix}\\
\hline
                 & v1                      &v2                     & v3                          & v4                   & v5                       & v6      &  v7            & v8&v9 \\ \hline
danceability     & 0.36                    & 0.66                    & -0.29                   & 0.29                    & -0.05                   & 0.27                    & 0.15                    & -0.08                   & 0.19                    \\ \hline
energy           & 0.85                    & -0.03                   & 0.27                    & -0.15                   & 0.02                    & 0.15                    & -0.1                    & 0.2                     & -0.2                    \\ \hline
valence          & 0.39                    & 0.77                    & 0.07                    & -0.14                   & 0.02                    & 0.29                    & 0.02                    & 0.02                    & 0.14                    \\ \hline
tempo            & 0.28                    & -0.03                   & 0.33                    & -0.65                   & 0.13                    & -0.16                   & 0.43                    & -0.32                   & 0.24                    \\ \hline
loudness         & 0.83                    & 0.01                    & 0.11                    & -0.1                    & -0.01                   & -0.03                   & -0.22                   & 0.06                    & -0.23                   \\ \hline
key              & 0.04                    & 0.02                    & 0.03                    & 0.2                     & 0.97                    & -0.03                   & -0.08                   & -0.02                   & 0.03                    \\ \hline
acousticness     & -0.82                   & 0.15                    & -0.04                   & 0.11                    & -0.03                   & -0.17                   & 0.06                    & -0.1                    & 0.13                    \\ \hline
instrumentalness & -0.42                   & -0.26                   & 0.01                    & -0.17                   & 0.1                     & 0.46                    & 0.35                    & 0.6                     & 0.1                     \\ \hline
liveness         & 0.06                    & -0.12                   & 0.73                    & 0.26                    & -0.11                   & 0.01                    & -0.33                   & 0.1                     & 0.5                     \\ \hline
speechiness      & 0.11                    & 0.09                    & 0.54                    & 0.52                    & -0.06                   & -0.12                   & 0.56                    & -0.02                   & -0.28                   \\ \hline
duration\_ms     & 0.01                    & -0.49                   & 0.04                    & 0.13                    & -0.01                   & 0.69                    & -0.01                   & -0.51                   & -0.04                   \\ \hline
popularity       & 0.69                    & -0.32                   & -0.37                   & 0.19                    & -0.03                   & -0.15                   & 0.16                    & 0.02                    & 0.28                    \\ \hline
year             & 0.74                    & -0.37                   & -0.29                   & 0.18                    & -0.02                   & -0.13                   & 0.17                    & 0.07                    & 0.18                    \\ \hline
\end{longtable}
\end{center}

To eliminate the effect of dimension, it needs to standardize all variables. The transformation equation is as follows.

\begin{equation}\label{8}
Z=\frac{X-\mathrm{E}\left( X \right)}{\sqrt{\mathrm{D}\left( X \right)}}
\end{equation} 

\vspace{0.2cm}

\subsubsection{Construction of measure index}

After performing Principal Component Analysis, the dimensionality is still high, so the traditional Euclidean distance, and Cosine similarity \cite{salton1988term} cannot be used. While facing high dimensional data, Heidarian \cite{heidarian2016hybrid} et al. constructed a metric applicable to extensive document similarity.

\begin{itemize}
\item \textbf{Triangle’s Area Similarity (TS)}
\end{itemize}

TS calculates the area of the triangle between two vectors and uses it as a similarity metric. In fact, three characteristics have been included in computing the similarity of vectors, namely the angle between vectors, the magnitude of vectors, and ED \cite{heidarian2016hybrid}.

To calculate the TS between vector $\boldsymbol{A}$ and vector $\boldsymbol{B}$, it is given by:

\vspace{0.2cm}

\begin{equation}\label{9}
\mathrm{TS}\left( \boldsymbol{A},\boldsymbol{B} \right) =\frac{\left| \boldsymbol{A} \right|\cdot \left| \boldsymbol{B} \right|\cdot \sin \left( \theta ^{\prime} \right)}{2}
\end{equation}

\vspace{0.2cm}

\begin{equation}\label{10}
|\boldsymbol{A}|=\sqrt{\sum_{n=1}^k{\boldsymbol{A}_{\boldsymbol{n}}^{2}}},\ \ \ |\boldsymbol{B}|=\sqrt{\sum_{n=1}^k{\boldsymbol{B}_{\boldsymbol{n}}^{2}}},\ \ \ \ \theta ^{\prime}=\cos ^{-1}\left( \frac{\sum_{n=1}^k{\boldsymbol{A}_{\boldsymbol{n}}}\cdot \boldsymbol{B}_{\boldsymbol{n}}}{|\boldsymbol{A}|\cdot |\boldsymbol{B}|} \right) +10
\end{equation}
Where the term $+10$ in $\theta ^{\prime}$ is because when vectors are overlapped, we cannot form the triangle. Thus, we increased theta by 10 degrees, keeping it rounded and avoiding more decimal points in calculations.

\newpage

\begin{itemize}
\item \textbf{Sector’s Area Similarity (SS)}
\end{itemize}

\begin{figure}[!h]
\centering
\includegraphics[width=6.5cm]{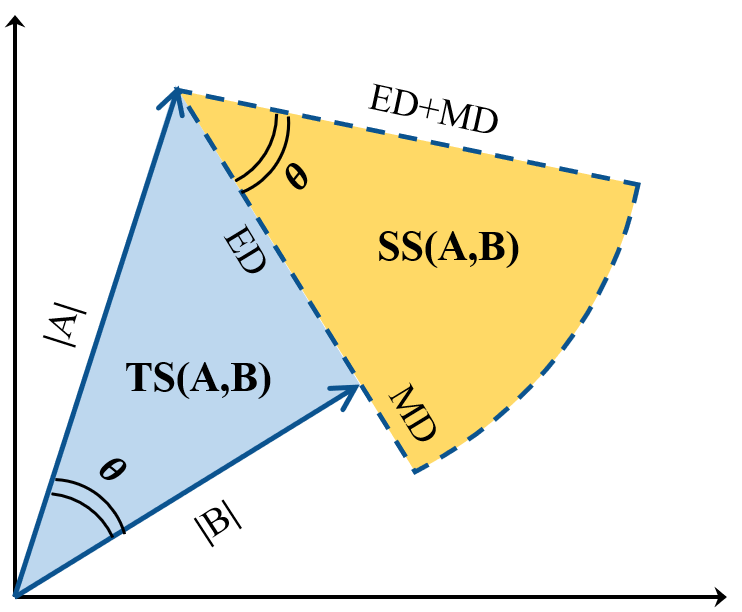}
\caption{Schematic diagram of component relationship}
  \label{fig:22-2}
\end{figure}

However, only relying on TS is not robust enough to interpolate vectors’ differentiations precisely to produce accurate similarity. As shown in Figure ~\ref{fig:22-2}, a new component should be added to describe the difference between the vectors’ Magnitudes.

Euclidean distance (ED), one of the most commonly used similarity metrics, constructs the metric positively correlated with MD as follows:

\begin{equation}\label{11}
\mathrm{MD}\left( \boldsymbol{A},\boldsymbol{B} \right) =\left| \sqrt{\sum_{n=1}^k{\boldsymbol{A}_{\boldsymbol{n}}^{2}}}-\sqrt{\sum_{n=1}^k{\boldsymbol{B}_{\boldsymbol{n}}^{2}}} \right|
\end{equation}

ED is also directly affected by the vector angle, which is the essential similarity measure. To enhance the ability to describe similarity, ED and MD are combined to form a radius of a circle, and then we obtain a sector with the angle of the vectors. Thus, the SS is calculated as follows.
\begin{equation}\label{11}
\mathrm{SS}\left( \boldsymbol{A},\boldsymbol{B} \right) =\pi \cdot \left(\mathrm{ ED}\left( \boldsymbol{A},\boldsymbol{B} \right) +\mathrm{MD}\left( \boldsymbol{A},\boldsymbol{B} \right) \right) ^2\cdot \left( \frac{\theta ^{\prime}}{360} \right) 
\end{equation}

\begin{equation}\label{11}
\mathrm{ED}\left( \boldsymbol{A},\boldsymbol{B} \right) =\sqrt{\sum_{n=1}^k{\left( \boldsymbol{A}_{\boldsymbol{n}}-\boldsymbol{B}_{\boldsymbol{n}} \right) ^2}}
\end{equation}

\begin{itemize}
\item \textbf{Similarity Measure Index (TSS)}
\end{itemize}

The purpose of constructing the Similarity Measure Index (TSS) is further strengthening the similarity metric. With occasionally the pretty difference between the magnitudes of TS and SS, the product is proposed to reduce the circumstance but still maintains the zero as TS or SS equaling to zero. Therefore, the final metric TSS is calculated as follows:
\begin{equation}\label{11}
\mathrm{TSS}\left( \boldsymbol{A},\boldsymbol{B} \right) \ = \ \mathrm{TS}\left( \boldsymbol{A},\boldsymbol{B} \right) \cdot \mathrm{SS}\left( \boldsymbol{A},\boldsymbol{B} \right) 
\end{equation}

So now the similarity of two songs (two vectors) can be measured, but note that a larger TSS indicates a lower similarity, and a TSS equal to 0 means identical.

\subsubsection{Test for similarity measure index}

\vspace{0.3cm}

One of the similarity metrics is Uniqueness. The three similarities (Euclidean distance, Cosine similarity, and TSS) of 50, 100, and 500 songs to each other were calculated separately, and their Uniqueness was counted. The results are shown in Table 2.4. (the precision of the calculation was determined by keeping seven decimal places).

\begin{center}
\begin{longtable}{cccc}
\caption{Uniqueness result table}\\
\hline
Number   of songs & Euclidean & Cosine   & TSS      \\ \hline
50                & 88.462\%  & 12.461\% & 98.116\% \\ \hline
100               & 92.148\%  & 16.847\% & 99.324\% \\ \hline
500               & 98.157\%  & 8.635\%  & 99.998\% \\ \hline
\end{longtable}
\end{center}

\vspace{0.3cm}

It can be seen that TSS have better Uniqueness on large datasets. Although Euclidean distance also has higher Uniqueness, the structural information can reflect is necessarily limited. Simultaneously, Cosine is limited by the computational accuracy, which leads to its poor performance on large datasets (in fact, when the computational accuracy is very high, we can already consider that the two are exactly similar, so the limitation on the computational accuracy is made).

After that, we randomly selected a song \textit{Barbara Lynn, 1963,You'll Lose A Good Thing}, use similarity calculation to get the most similar song \textit{Carly Simon,1974,Haven't Got Time for the Pain}\footnote{An interesting phenomenon is that the song was re-released in 1975 with the same characteristics}. We searched and listened to two songs, which matched the musical similarity.

In summary, TSS has better similarity metric performance from a Uniqueness perspective. We will apply our metrics in the next section to determine whether artists are more similar within genres than between genres.

\vspace{0.3cm}
\subsection{Genre Analysis Model}

\vspace{0.3cm}
\subsubsection{Comparative Analysis of Similarities within and outside the Genre}

Traversing all artists is an excellent choice to perform similarity analysis within and outside genres. Without considering the complexity of computing the metrics, the time complexity of the traversal is $O\left( n^2 \right) $. Simple random sampling is taken to compute the similarity within and outside of genres to reduce the computational effort.

For this large dataset, 2500 data per sample is sufficient to reflect the overall level. 
To calculate the similarity within the genre $SWG$, it needs to extract an artist $q$ of genre $m$, and another artist $p$ of the same genre, and then calculate the similarity $TSS\left( \boldsymbol{A}_{mq},\boldsymbol{B}_{mp} \right)$.
While for the similarity outside the genre $SBG$, it needs to extract an artist $q$ of genre $m$, and artist $p$ from any other genre $n$, and then calculate the similarity $TSS\left( \boldsymbol{A}_{mq},\boldsymbol{C}_{np} \right)$.
Finally, the average similarity is compared to indicate the strength of similarity within and outside the genre. The average similarity can be defined as follows.
\begin{equation}\label{15}
SWG=\sum_{i=1}^N{\sum_{j=1}^{2500}{TSS\left( \boldsymbol{A}_{mq},\boldsymbol{B}_{mp} \right)}}
\end{equation}
\begin{equation}\label{16}
SBG=\sum_{i=1}^N{\sum_{j=1}^{2500}{TSS\left( \boldsymbol{A}_{mq},\boldsymbol{C}_{np} \right)}}
\end{equation}
Among them, $m,n,p,q$ are the sample data subscript.

\begin{figure}[htbp]
\centering    
 \subfigure[Sampling results within genres] 
{
	\begin{minipage}{7cm}
	\centering          
	\includegraphics[scale=0.56]{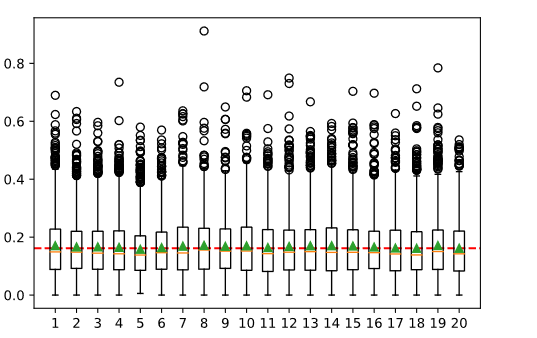}   
	\end{minipage}
}
	\subfigure[Sampling results between genres] 
{
	\begin{minipage}{7.3cm}
	\centering      
	\includegraphics[scale=0.56]{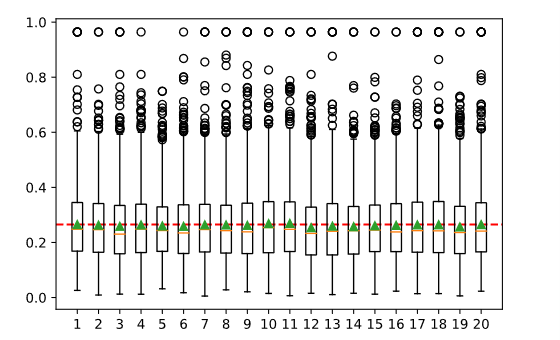}   
	\end{minipage}
}
 \caption{Sampling results} 
\label{fig:33}  
\end{figure}

After that, 20 simple random samples were executed, and the results showes a stronger similarity within genres than between genres. The sampling results are shown in Figure ~\ref{fig:33}.


\vspace{0.1cm}
\subsubsection{Comparative Analysis of the Influence within and outside the Genre}
\vspace{0.1cm}
If there is a connection between two network nodes (artists), it can assume that there is an influence between the two artists. The node influence metric $NI_i$ obtained from the previous model can be utilized to measure this influence between nodes. Then the influence between node $q$ from genre $m$ and node $p$ from genre $n$ can be described by the normalized rank of the nodes' influence. 

\begin{equation}\label{17}
IP_{qp}^{mn}=\frac{1}{1+|\mathrm{Rank}\left( NI_{q}^{m} \right) -\mathrm{Rank}\left( NI_{p}^{n} \right) |}
\end{equation}

\vspace{0.3cm}

Then it goes back to the “Comparative Analysis of Similarities” problem. It can also use simple random sampling for processing. The formula for calculating influence $WIP$ within genres and influence $TIP$ between genres defined as follows: 

\begin{equation}\label{17}
WIP=\sum_{i=1}^N{\sum_{j=1}^{2500}{IP_{qp}^{mm}}}
\end{equation}

\begin{equation}\label{17}
TIP=\sum_{i=1}^N{\sum_{j=1}^{2500}{IP_{qp}^{mn}}}
\end{equation}

\vspace{0.2cm}

\begin{figure}[h!]
\centering 
\subfigure[Sampling results within genres] 
{
\begin{minipage}{7cm}
\centering 
\includegraphics[scale=0.56]{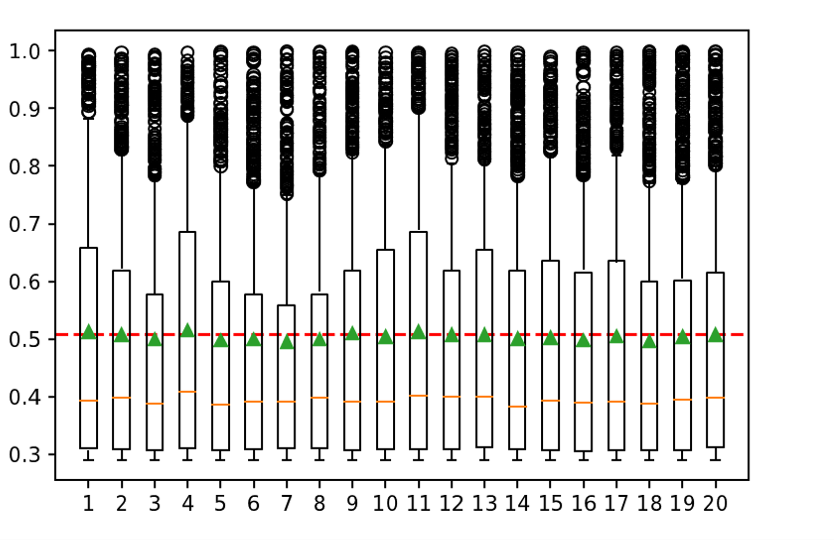} 
\end{minipage}
}
\subfigure[Sampling results between genres] 
{
\begin{minipage}{7.3cm}
\centering 
\includegraphics[scale=0.56]{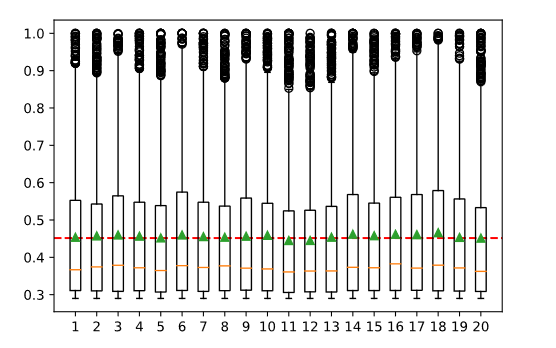} 
\end{minipage}
}
\caption{Sampling results} 
\label{fig:333} 
\end{figure}

After that, 20 simple random samples were executed. The results showes that there was more substantial similarity within genres than between genres, which are shown in Figure ~\ref{fig:333}.

\subsubsection{Universality and Individuality of the Genre}
In this subsection, the interrelationship between genres will be reflected through Cluster analysis and Time series analysis, with the aim of studying the commonalities and individuality of genres.

First, a systematic clustering of twenty genres is performed, the clustering results are as 
follows.

\begin{figure}[h!]
\centering
\includegraphics[width=11.7cm]{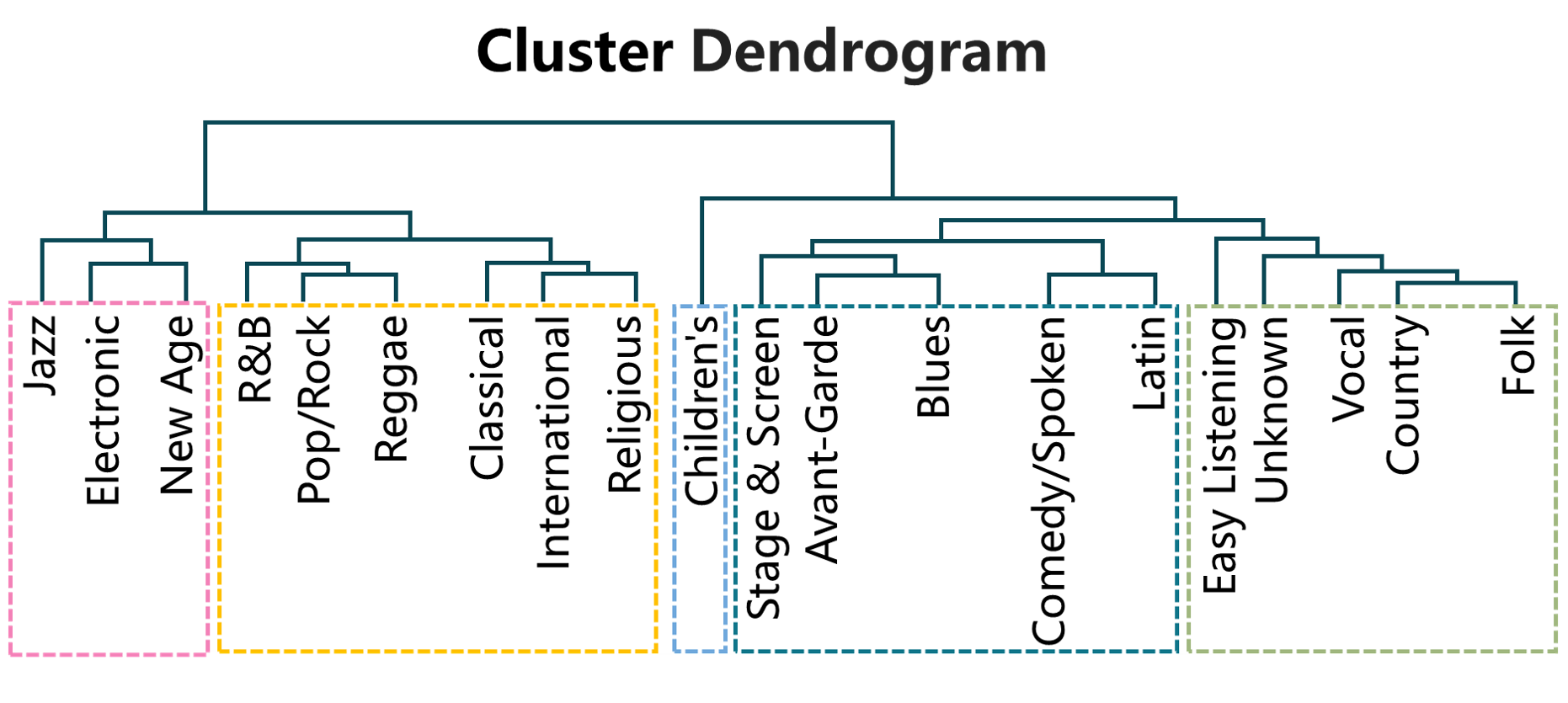}
\caption{Clustering result graph}
\label{fig:33-5}
\end{figure}

This allows us to further explore the differences between genres.

\begin{enumerate}[\bfseries 1.]
\item Children's is a separate category, probably because of the vast difference in audience, which leads to a considerable difference from other genres \cite{marsh2008musical}.
\item Easy Listening, Vocal, Country, and Folk are one category. The common feature is that the Tunes are more soothing and use fewer instruments, i.e., less Energy and Instrumentation \cite{hu2007exploring}.
\item Jazz, Electronic, and New Age are in one category. They use more instruments and prefer to use electronic sounds or instrumental sounds instead of human voices, i.e., lower Acousticness and higher Instrumentalness \cite{schrank1999big}.
\item Pop/Rock, R\&B, Reggae, Classical, International, Religious are divided into one category. Probably they all have some influence on each other and finally pass through the network, making these six genres more similar. Pop/Rock, R\&B, Reggae have higher Energy, Valence, and Loudness, reflecting popular elements. Classical, International, and Religious have lower Valence, which also shows the characteristics of traditional music \cite{schindler2012facilitating}.
\item Stage \& Screen, Unknown, Avant-Garde, Comedy/Spoken, Latin are one category. Comedy/Spoken, Stage \& Screen have higher Speechiness, while Latin has higher Danceability and is more suitable as a dance music accompaniment \cite{schindler2012facilitating}.
\end{enumerate}

\begin{figure}[!h]
\centering
\includegraphics[width=11.5cm]{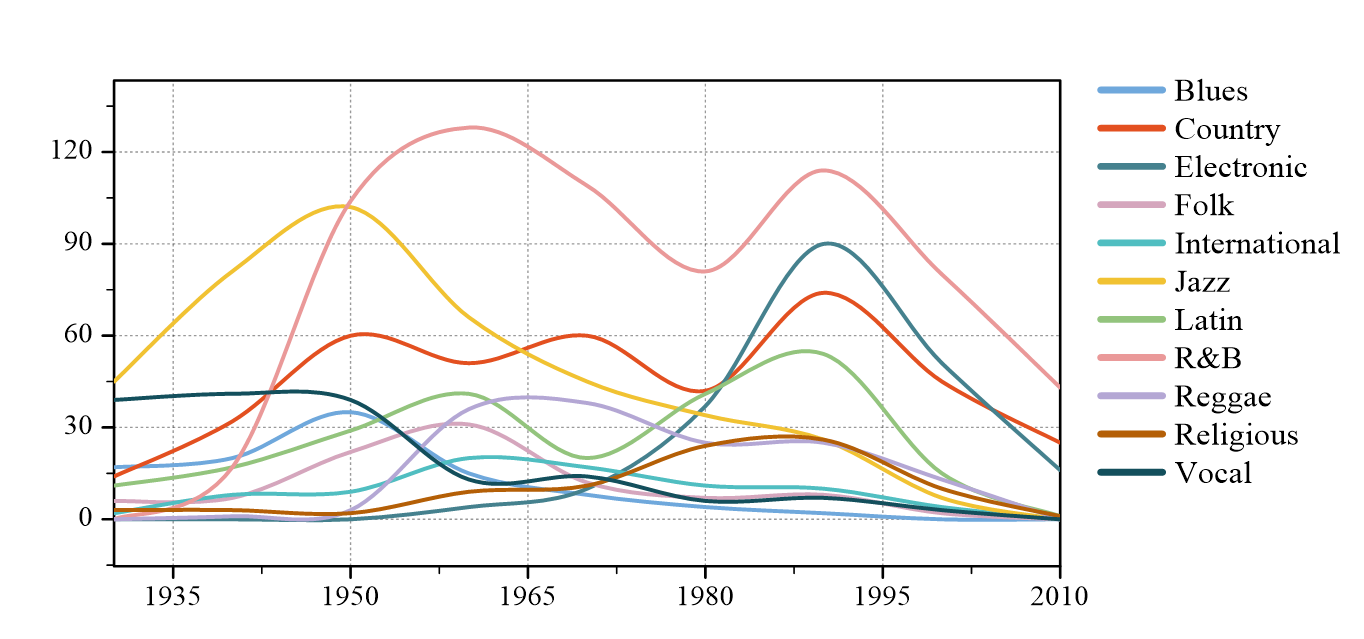}
\caption{The change in the number of genres to debut over time}
\label{fig:44-1}
\end{figure}

After further processing the data, it obtained the number of debutants per genre per year. Due to a large amount of data, it is sufficient to select a representative interval and draw a line graph to visualize the change of the number of debutants of each genre over time, as shown in Figure ~\ref{fig:44-1}.

\vspace{-0.5cm}

\begin{figure}[!h]
\centering
\includegraphics[width=13.5cm]{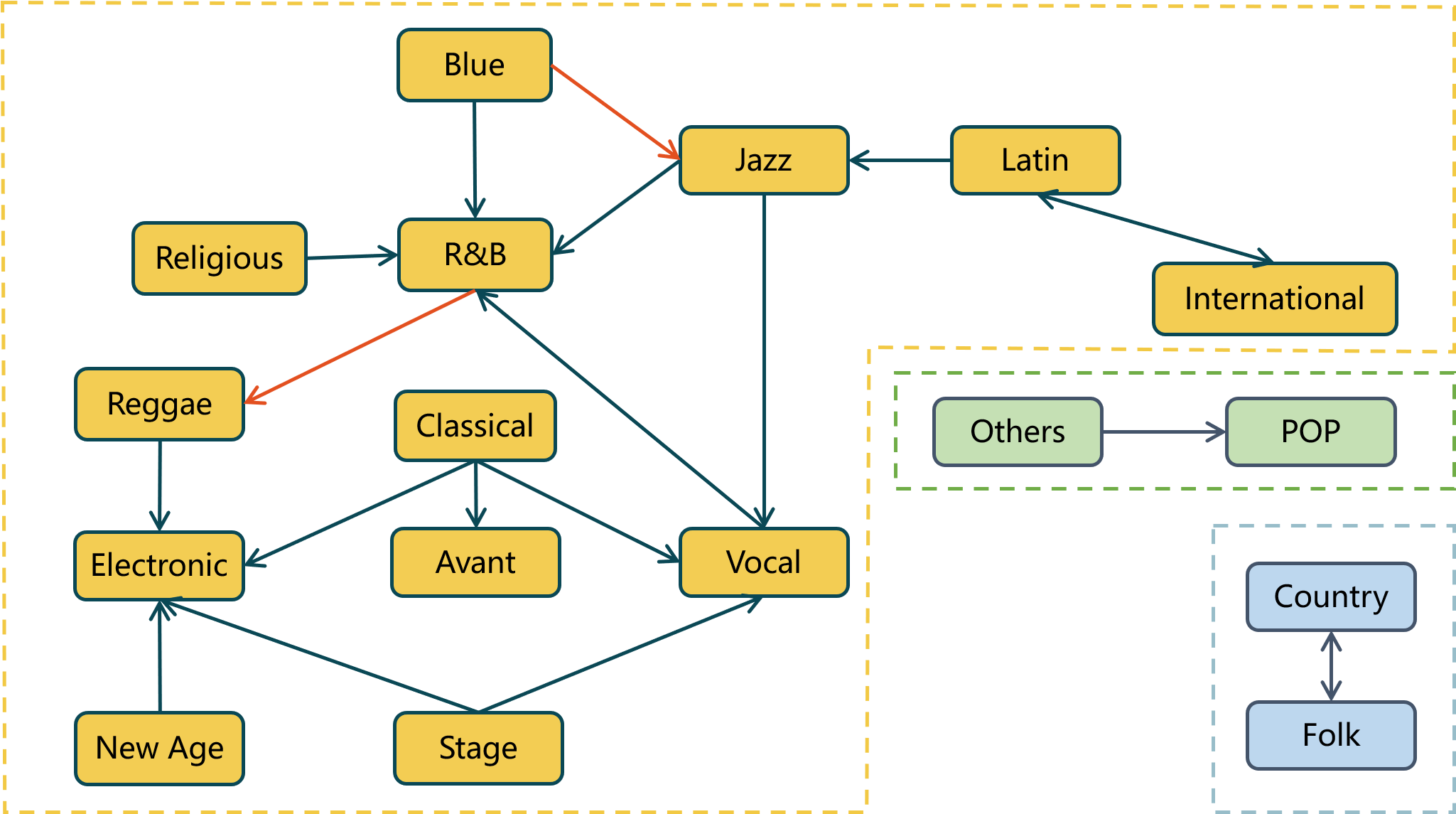}
\caption{Influence relationship between genres.}
\label{fig:44-2}
\end{figure}

By processing the data at a deeper level and eliminating the minor influences, we can obtain the influence of each genre on the other genres and determine the direction of influence. Then connecting the influences of each genre, as shown in Figure ~\ref{fig:44-2}.

\begin{figure}[htbp]
\centering    
 \subfigure[Special case 1] 
{
	\begin{minipage}{7.2cm}
	\centering          
	\includegraphics[scale=0.5]{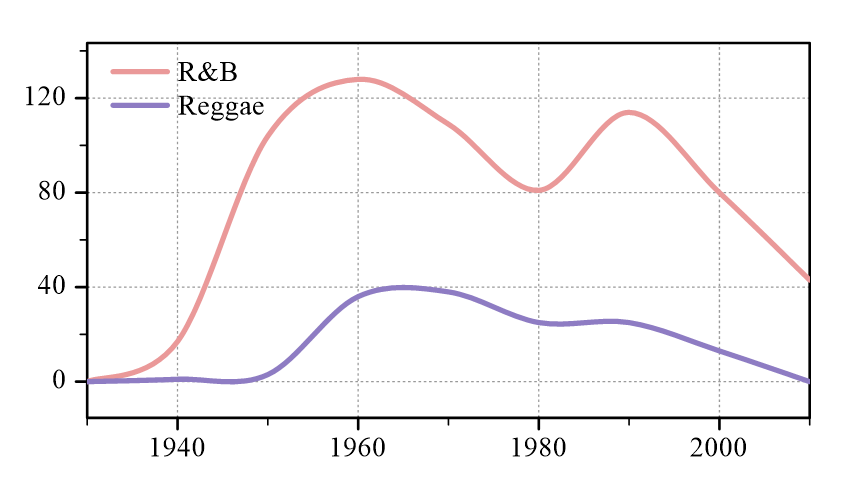}   
	\end{minipage}
}
	\subfigure[Special case 2] 
{
	\begin{minipage}{7.1cm}
	\centering      
	\includegraphics[scale=0.5]{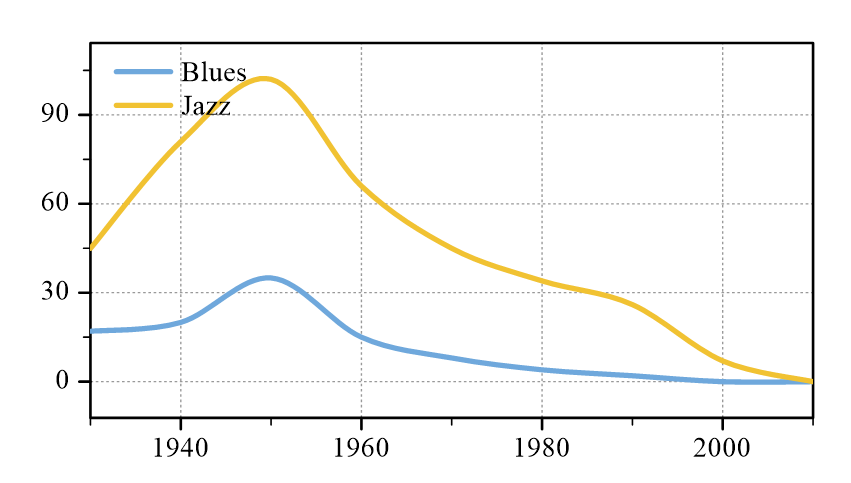}   
	\end{minipage}
}
 \caption{The two exceptional cases marked with red arrows in Figure ~\ref{fig:44-2}} 
\label{fig:44}  
\end{figure}

Nevertheless, there may be stronger influences among the removed secondary influences. In order to check out these particular cases, it needs to combine the trends of each genre over time. After that, we found two exceptional cases, marked with red arrows in Figure ~\ref{fig:44}.

\vspace{-0.5cm}

\vspace{-0.3cm}

\subsection{Authenticity of Influence and the Revolutionaries in Influence}
\subsubsection{Authenticity of Influence}

\begin{figure}[!h]
\centering
\includegraphics[width=11.7cm]{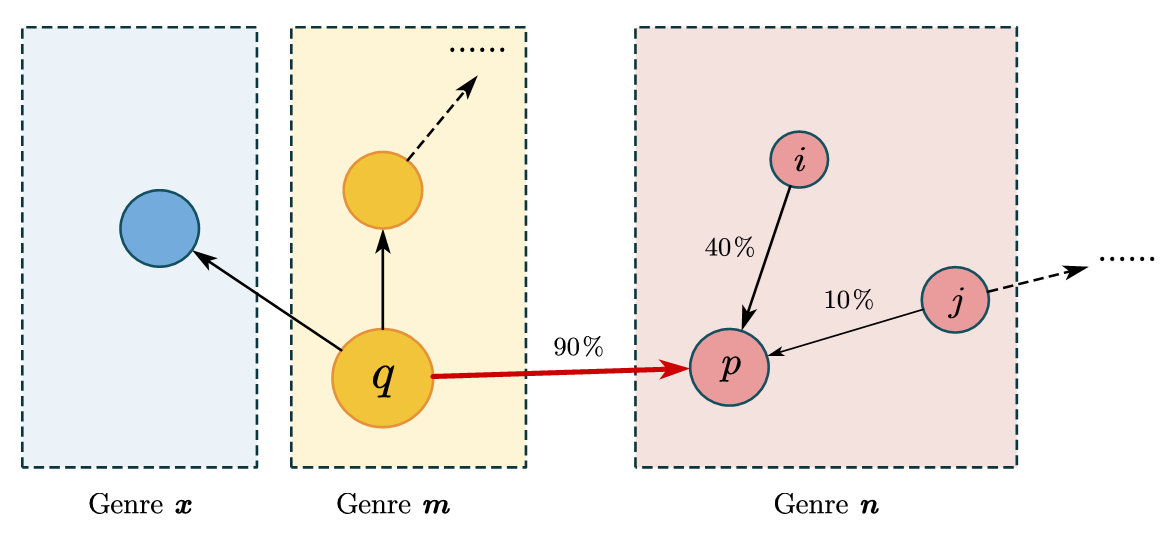}
\caption{Schematic diagram of the real significant impact}
\label{fig:55-1}
\end{figure}

First, a significant situation was noted. As shown in Figure ~\ref{fig:55-1}, an artist $q$ of genre $m$ influences an artist$ p $of genre $n$, and the similarity between these two artists is as high as 90\%. While two other artists $i,j $ of genre $n$ also influence artist $p$ of the same genre, but their similarity is low. In such a case, we admit that artist $q$ truly influences artist $p$. If the distribution of similarity between all followers (pointed nodes) and his influencers (nodes) in a music network shows the above significant situation, then it can be considered that the similarity metric reflects the true nature of this influence. We refer to this similarity distribution as “extreme distribution.”

Based on such analysis, we counted the similarity distribution of each node in the network. Those nodes with only one input edge were temporarily discarded (632 nodes), and we wanted as many nodes with high and low similarity as possible. For these values mapped to the $[0,1]$, we define the average zero-one distance $AD_i$ of a network node $i$. The average Euclidean Distance of the distribution between this node is calculated, and the closer this distance is to 1, the closer it is to our significant state, as follows.

\begin{equation}\label{20}
AD_i=\frac{2}{n\left( n-2 \right)}\sum_{m\ne n}{|TSS_m-TSS_n|},\ \ m,n\in i
\end{equation}

\begin{figure}[!h]
\centering
\includegraphics[width=15.3cm]{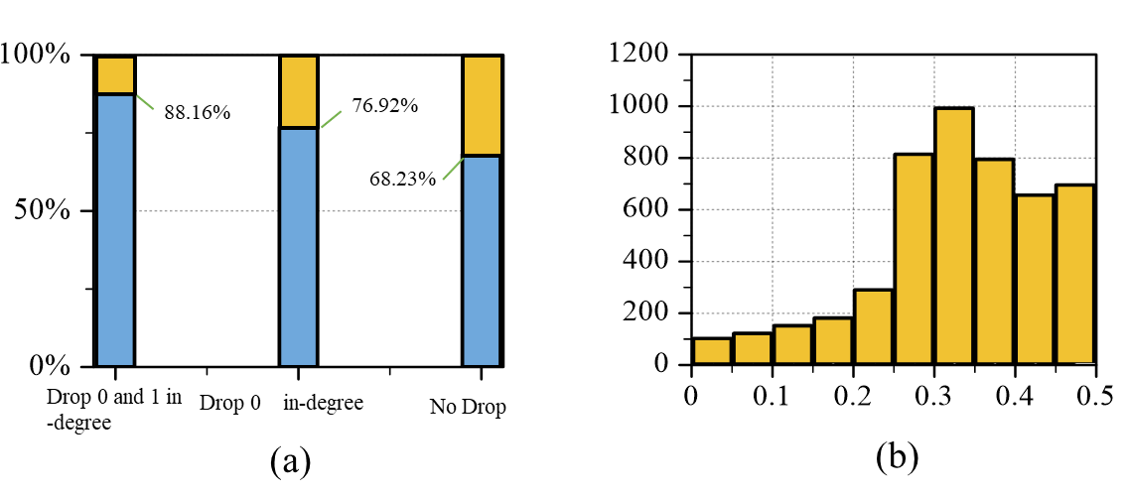}
\caption{Standard deviation of the similarity distribution of each network node }
\label{fig:55-2}
\end{figure}

A threshold $\alpha =0.8$ was set, and results above this threshold value are considered to fit the “extreme distribution.” The calculation results are shown in Figure ~\ref{fig:55-2}(a) below. We find that with the exclusion of nodes with no input edges and only one input edge (4324 nodes), 88.16\% \footnote{If it is not ruled out that there is only one input edge, it is 76.92\%} of the data follow a significant zero-one distribution.

To further corroborate our conclusion, the standard deviations of the similarity distributions of individual network nodes were calculated and a distribution plot of these standard deviations was made, as shown in  Figure ~\ref{fig:55-2}(b).For the values mapped to $[0,1]$, standard deviations can roughly indicate that many similarity distributions are biased toward “extreme distributions.” Therefore, we believe that the similarity data can suggest that the identified influencers in fact influence the respective artists.

Since the similarity metric calculates the similarity between two artists, and these artists' data are calculated from the mean value of their songs, does this allow us to logically suggest that 'influencers' actually affect the music created by the followers? What can be noticed is that some songs have multiple artists co-creating them, whether there exists the possibility that multiple artists' genres are different. When retrieving this situation, it finds a situation where part of the Artist ID in this dataset does not appear in any other dataset. This situation described above disappeared when these data were cleaned out. Therefore we cannot give further analysis here. 

Having established that influencers do influence the music created by their followers, it needs to further explore whether their influence is related to some “infectious” musical characteristics.

To explore whether the influence is related to musical characteristics, the most straightforward idea is to perform multiple linear regression. However, our influence metric does not involve any music features, and involving unrelated variables would compromise the model's validity. Therefore the Lasso Regression \cite{ranstam2018lasso} is used, which can avoid this problem by L1-norm. However, since Lasso regression is not good at dealing with highly correlated variables, the Elastic Net Regression model \cite{hans2011elastic} was used, which can be defined as follows.

\vspace{0.2cm}

\begin{equation}\label{21}
\min_{\left( \beta _0,\beta \right) \in R\times R^{\rho}} \left\{ \frac{1}{2}\sum_{i=1}^N{\left( y_i-\beta _0-x_{i}^{T}\beta \right) ^2}+\lambda \left[ \frac{1}{2}\left( 1-\alpha \right) \left\| \beta \right\| _{2}^{2}+\alpha \left\| \beta \right\| _1 \right] \right\} 
\end{equation}

\vspace{0.4cm}

The coefficient matrix obtained from conducting the regression results is shown in Table 2.5, and these coefficients indicate the strength of musical characteristics on the influence of the artist.

\begin{center}
\begin{longtable}{cccccc}
\caption{Coefficient matrix table}\\
\hline
danceability & energy & valence & tempo & loudness & key \\ \hline
1.0501 & -1.5912 & 1.3140 & 0 & 0 & 0.0955 \\ \hline
acousticness & instrumentalness & liveness & speechiness & duration & popularity \\ \hline
-1.478 & -0.6981 & 2.982 & -4.2148 & 0 & -0.0423 \\ \hline
\end{longtable}
\end{center}

Speechiness has the most substantial influence, followed by Liveness; Valence, Energy, and Acousticness have a weaker but similar influence, followed by Danceability, Key, and Popularity in decreasing strength. In addition, Tempo, Loudness, and Duration have almost no influence.

\vspace{0.5cm}

\subsubsection{The Evolution of Music}

\vspace{0.3cm}

How revolution is reflected in the data is very difficult. In MIDN Model, The Beatles are considered the most influential musicians, and Wikipedia considers them the most influential band. 
If Wikipedia can label musicians as major “Revolutionaries/non-major Revolutionaries”, this translates into a common classification problem in machine learning. 
The models commonly used to deal with these classification problems deliver feature significance, which provides a good way to find features that can reveal major revolutions.

However, the problem now is that there is no God to label us, so we need to label the data. Based on the influence rankings obtained from the Influence model, the bottom 10\% of musicians are labeled as “ non-major revolutionaries” (0/1 label), which is easy to understand. For the top 20\% of musicians, we used Network analysis and Semantic analysis to decide whether to label them as “major revolutionaries”.

\begin{itemize}
\item \textbf{Network analysis:}
\end{itemize}

First, the network will be reconstructed, differing from the previous one, in that the size of the nodes is measured using the artist's influence metric and the edge weights of the nodes are the same as before. After that, we perform year clustering for each genre.

\begin{figure}[!h]
\centering
\includegraphics[width=13cm]{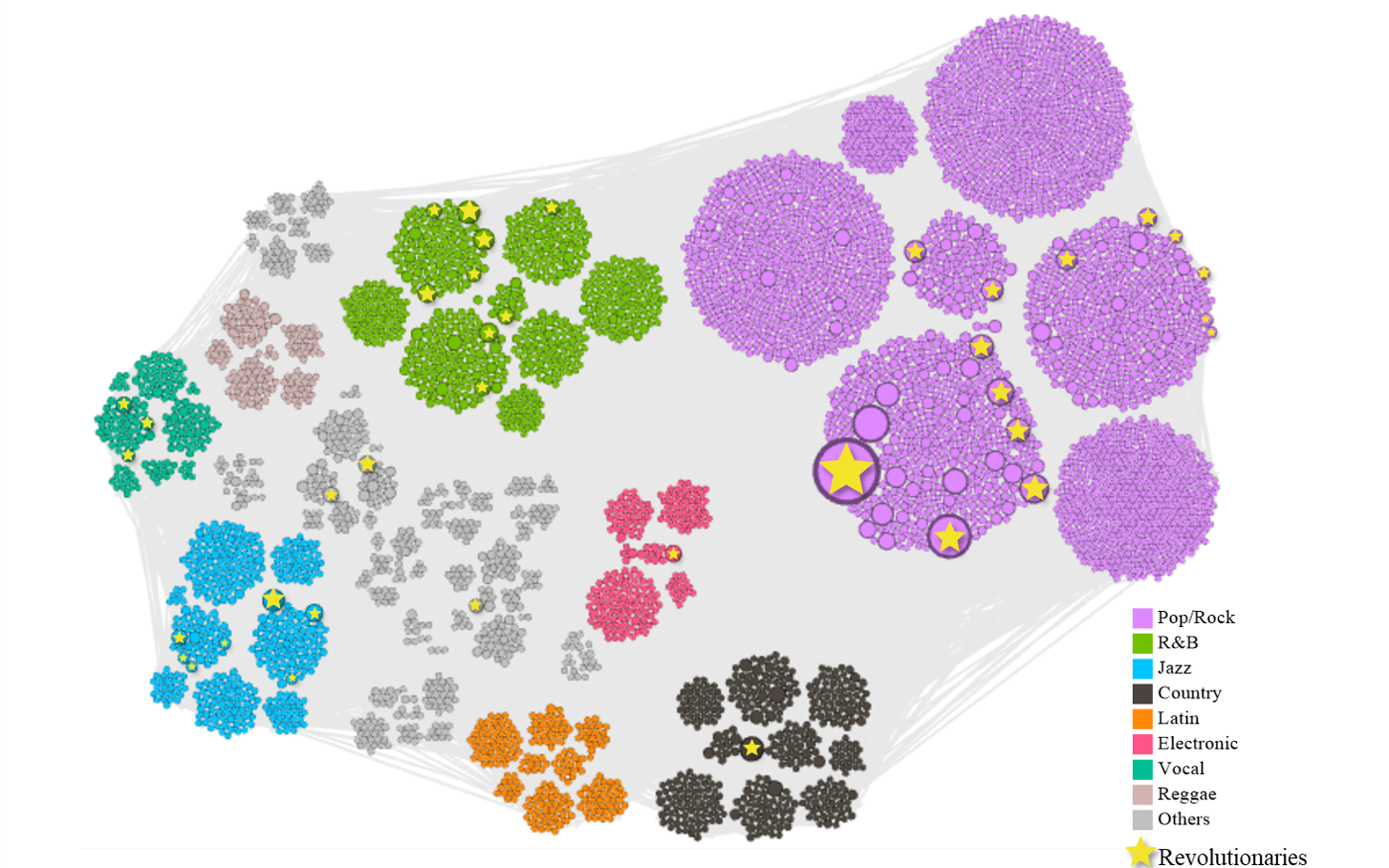}
\caption{Clustering network diagram and major revolutionaries}
\label{fig:66-1}
\end{figure}

To determine the relationship between the nodes' positions after clustering. The idea is proposed that the more influential an artist is to other genres, the more he should be outside the clustering group because he needs to keep a short distance from the artist in other genres. Based on such clustering, those revolutionaries should be on the periphery of the clustering group and have greater influence (larger nodes). The network after clustering and the major revolutionaries we found is shown in Figure ~\ref{fig:66-1}.

\newpage

\vspace{-0.1cm}

\begin{itemize}
\item \textbf{Semantic analysis:}
\end{itemize}

\vspace{-0.1cm}

First, Wikipedia words describing the status of a musician were collected to build a corpus that would indicate the musician as a revolutionary, including words such as “the most influential”, “major influence”, “Father of”, and other words.
Then we first determine if Wikipedia for each musician has the words present in the corpus. If it does not exist, then use Google again, using the search formula consisting of “musician” + “indicator word” to determine whether the first page of search results exists indicator word.

With the above analysis and the labeling, we can look for possible features through the Random Forest model \cite{rigatti2017random}, which, as a bagging model, coincides with the definition of major revolutionaries.

\begin{figure}[!h]
\centering
\includegraphics[width=10.5cm]{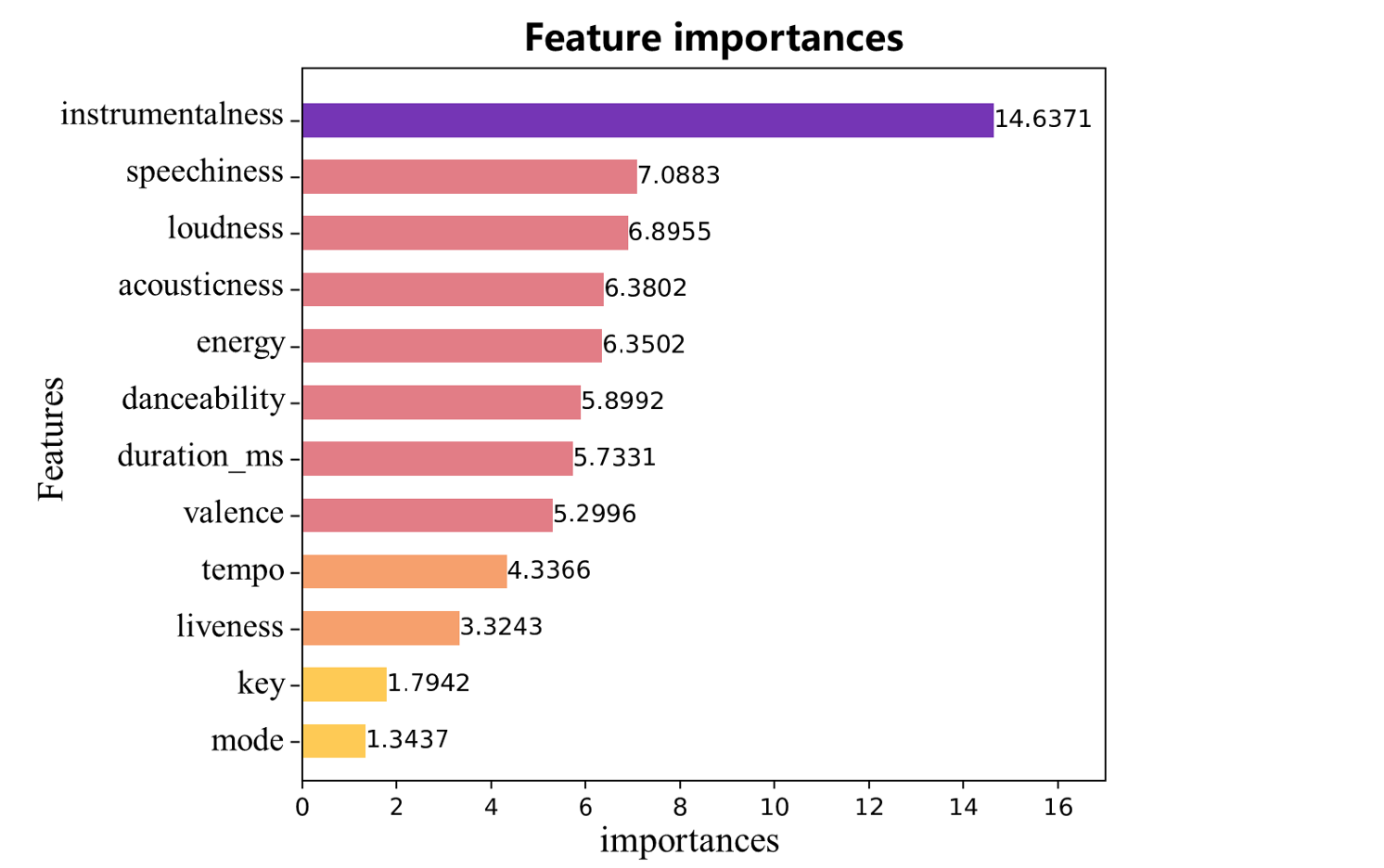}
\caption{Feature Importances}
\label{fig:66-2}
\end{figure}

Since the sample size of major revolutionaries is much smaller than that of non-major revolutionaries, we use the first 10\% of them as the training set, 10\%-15\% as the validation set, and 15\%-20\% as the test set. After training the model, feature importances are shown in Figure ~\ref{fig:66-2}.

It can be found that Instrumentalness is a potential feature that can indicate major revolutions. However, it was unable to test it in the remaining data because there were too few significant revolutionaries in the remaining data. Therefore, we can only consider this as a possible feature.

\section{Practical application and discussion of the model}

\subsection{List of data used in the model}

(1) “Music Influence Data”(These data were collected from the website AllMusic.com) represents musical influencers and followers, as reported by the artists themselves, as well as the opinions of industry experts. These data contains influencers and followers for 5,854 artists in the last 90 years. 

(2) “Full set of music data” (These data come from Spotify's API) provides 16 variable entries, including musical features such as danceability, tempo, loudness, and key, along with artist name and artist id for each of 98,340 songs. 

\subsection{The Evolution of Country Music}

The model will be practically applied to a genre to analyze the influence processes of musical evolution that occurred over time in one genre. To see if the proposed method can identify indicators that reveal the dynamic influencers and explain how the genre(s) or artist(s) changed over time? Specifically, Country music is chosen to study its evolutionary process.

\begin{figure}[!h]
\centering
\includegraphics[width=14.9cm]{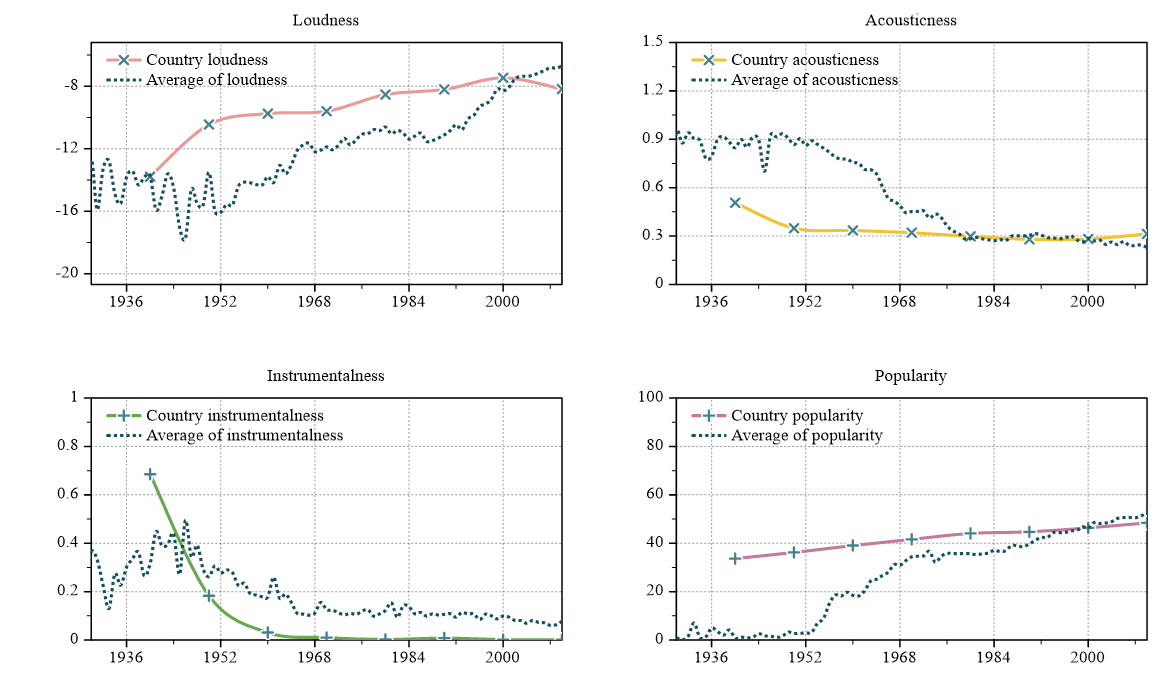}
\caption{Indicator trend diagram}
\label{fig:q6-1}
\end{figure}

First, it needs to correspond to the metrics of all artists over time with Country music. With the known data, the following four graphs were selected as control graphs, as shown in Figure ~\ref{fig:q6-1}.

The Loudness increased over time, i.e., the musical decibel level gradually increased. Instrumental has been decreasing and has remained at a low level, i.e., increasing the vocals in the music. This may be done by reducing the purely instrumental or changing from playing to singing along with the accompaniment. Popularity has been rising rapidly, probably because the public has become more interested in music. The difference between the two is also found: the average trend of Energy is increasing year by year, but the change in Country music is not much. At the same time, the duration of Country music is slowly decreasing, but the average trend is not much changed. The larger difference at the beginning indicates that the uniqueness of Country music is stronger in the earlier period. In aggregate, the changes in the indicators of Country music are similar to the average trend, indicating that country's evolution is influenced by the times \cite{fenster1993genre}.

In addition, it also needs to study the temporal distribution of the works produced and the changes in the number of people in this genre, as shown in Figure ~\ref{fig:q6-2} and Figure ~\ref{fig:q6-3} below:

\begin{figure}[htbp]
\centering
\begin{minipage}[t]{0.49\textwidth}
\centering
\includegraphics[width=7cm]{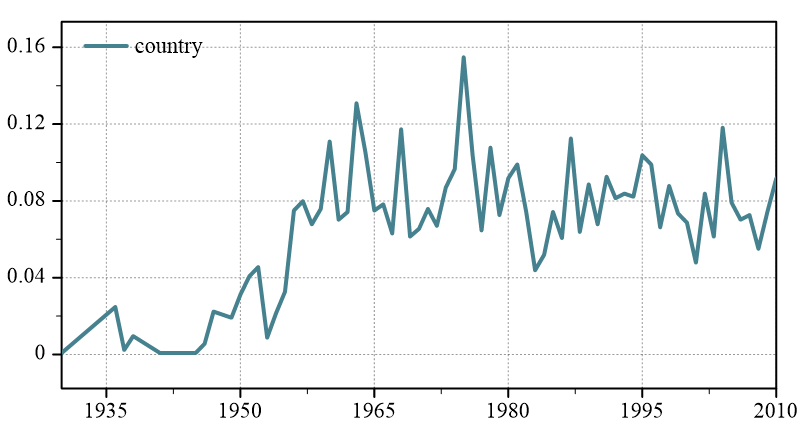}
\caption{Proportion of country music works }
\label{fig:q6-2}
\end{minipage}
\begin{minipage}[t]{0.49\textwidth}
\centering
\includegraphics[width=7cm]{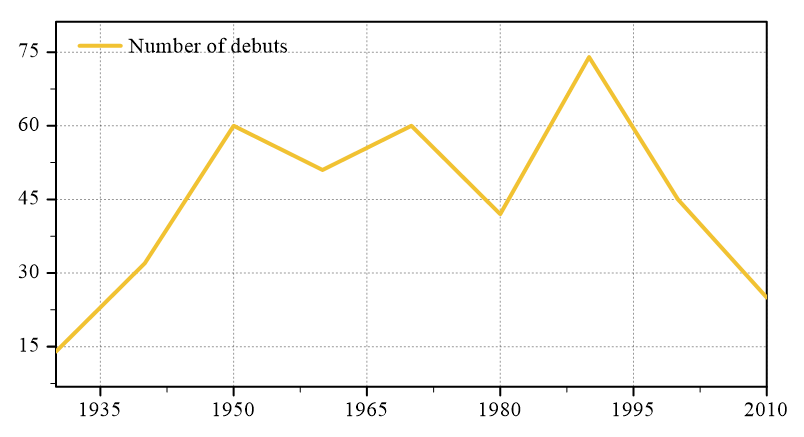}
\caption{ The number of debuts varies }
\label{fig:q6-3}
\end{minipage}
\end{figure}

It is evident that during World War II, country music's status in the popular entertainment industry increased dramatically, and its range of influence expanded tremendously. This was probably because many people at that time moved into urban factories and the military, placing their nostalgic feelings on Country music. 1950 to 1990 seemed to be steady growth in Country music.

\subsection{Environmental Impact on Music}

In this section, the proposed methods will be combined with the actual social environment to express the impact of music on culture.

\begin{figure}[htbp]
\centering    
 \subfigure[A] 
{
	\begin{minipage}{7cm}
	\centering          
	\includegraphics[scale=0.5]{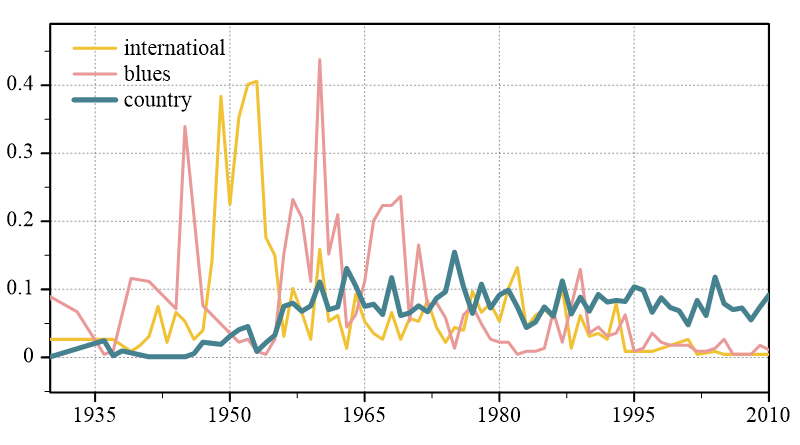}   
	\end{minipage}
}
	\subfigure[B] 
{
	\begin{minipage}{7cm}
	\centering      
	\includegraphics[scale=0.51]{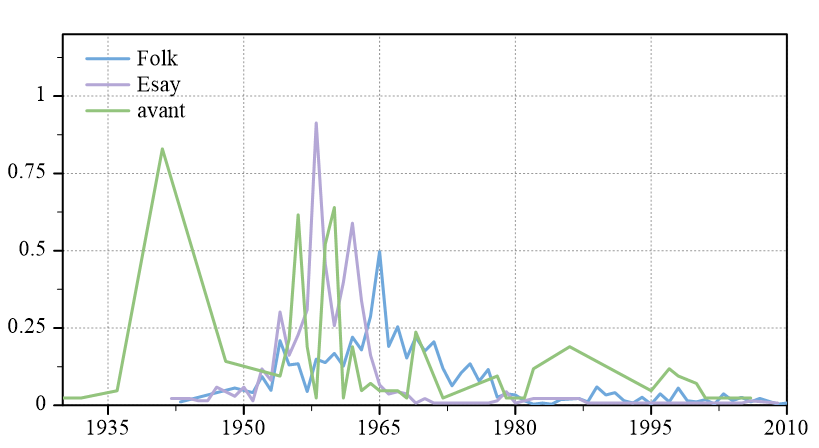}   
	\end{minipage}
}
 \caption{Changes in the proportion of music } 
\label{fig:q7-1}  
\end{figure}

\begin{figure}[htbp]
\centering    
 \subfigure[A] 
{
	\begin{minipage}{7.2cm}
	\centering          
	\includegraphics[scale=0.5]{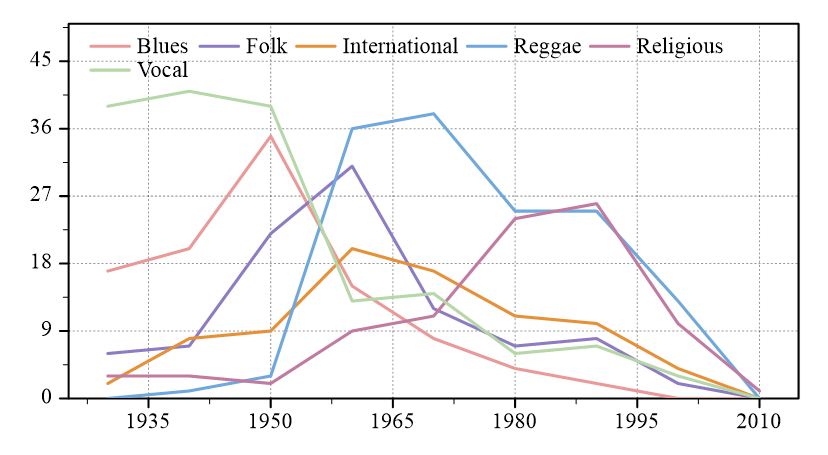}   
	\end{minipage}
}
	\subfigure[B] 
{
	\begin{minipage}{7.2cm}
	\centering      
	\includegraphics[scale=0.5]{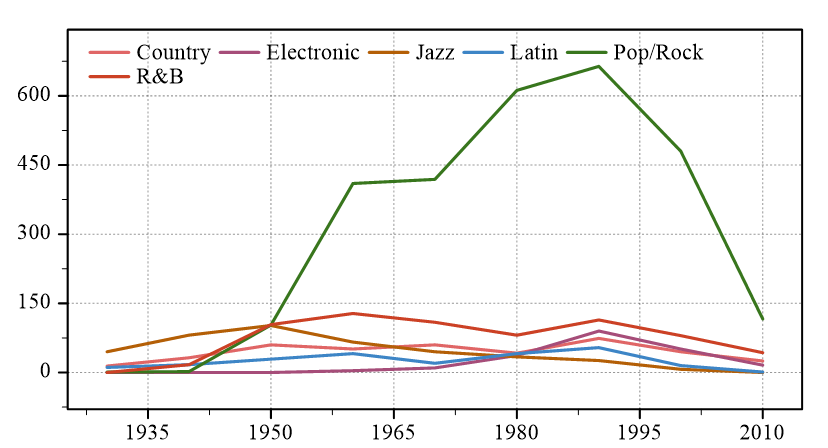}   
	\end{minipage}
}
 \caption{Changes in the number of debuts} 
\label{fig:q7-2}  
\end{figure}

The distribution of work production and the change in the number of debuts over time were mainly analyzed, as shown in Figure~\ref{fig:q7-1} and Figure ~\ref{fig:q7-2}.

\vspace{0.5cm}

Combining the above chart with Wikipedia's introduction, one can learn that.

\begin{enumerate}[\bfseries 1.]
\item After the late 1960s, the Electronic genre gradually became more and more popular because of technological innovations, and the use of electronic means to produce music became more convenient and inexpensive \cite{collins2013electronic}.
\item Avant-Garde maintained a high level of activity until 1960, but the fusion and development of various genres after the 1970s led to its gradual withdrawal from the mainstream \cite{mcclary1989terminal}.
\item Blues also had a high activity level from the 1930s to the 1970s \cite{oliver1998story}. After the 1970s, the rapid economic development of the United States led to a greater preference for pop music such as Pop and R\&B., similar to Folk music and International music.
\item Pop/Rock has maintained a high and steady activity level since the 1960s \cite{regev2013pop}. Since the early 1960s, almost all pop has been influenced by the rock in one way or another. But it was not until the 1970s (the period when the first generation of rock fans had come of age) that pure pop took the stage at the right tempo, leading to Pop/Rock becoming mainstream (as a revolt against the early rock that dominated much of the 1960s)
\end{enumerate}

The analysis is consistent with the network we constructed before, proving that the metrics and network we constructed can be integrated with the actual social environment.

\subsection{Discussion}

The approach is to discover the value of music by building music networks. When used to analyze music impact, conventional topology networks will only be based on some existing topology metrics. The influence metrics we construct take into account the propagation nature of music and the existing topology metrics and thus have a better performance in subnetwork analysis. What also needs to be considered is whether the data collected is a true reflection of the music's impact. The statistical results of this study show that the similarity metrics constructed can indicate the authenticity of the impact to some degree. Finally, the approach employed allows a deeper insight into the process of revolution in and development of music and may reveal some features that indicate the occurrence of revolution. This is valuable for discovering the understudied history of music.

In addition, the Music Influence Directed Network Model((MIDN Model) constructed reveals a complex relationship of mutual influence between genres. The differences between genres have gradually become smaller, and the evolution of most genres is similar to the evolution of the music world as a whole. This has demonstrated that different musical genres will merge, and musicians will constantly learn from others and suggested that the historical background may influence the development of music. However, there are also exceptions, i.e., Pop/Rock replaced early rock dominating much of the 1960s and became the dominant musical genre. This complex relationship can also be discovered by the methods given in this paper.

However, the model also has shortcomings. In the IAMR Model, we investigate whether the data can reliably reflect reality with the help of influence and similarity metrics. We propose the concept of “extreme distribution” and define the distance of a network node. However, because data of some of the songs' co-writers is missing and has to be deleted completely, it is impossible to conclude with certainty that “influencers” actually influence the music created by their followers. Therefore, it is important to consider whether our model to examine the authenticity of the influence should be further improved or whether the authenticity of the influence can be concluded through other methods. Second, in the process of influence metric construction, this paper constructs the function $f\left( x \right) =e^x-1$, which satisfies $x\longrightarrow \infty , y\longrightarrow 0, f\left( x \right) y\rightarrow \infty $ , but the ideal situation is $f\left( x \right) \cdot y\rightarrow x$. Since it is difficult to construct such a function, we weaken the condition to turn to $f\left( x \right) \cdot y\rightarrow \infty $. In this sense, it is also worth exploring whether a better function exists.

More research is needed in the field of music. Currently, people are increasingly aware of the copyright of music, yet music cultural appropriation is still a serious problem. Some plagiarism is disguised as drawing inspiration from others’ works, and such plagiarism techniques are becoming more sophisticated and difficult to detect. So, can we discover more underlying features to describe the similarity between music in a more accurate way? In addition, this study finds that the popularity of music and the changes of the times are positively correlated, so could music influence the cognitive process of the human brain, thus giving birth to a new musical genre that drives the development of music.

\section{Conclusion}

This paper aims to examine the evolution and revolutions of musical genres. This paper includes four models, namely Music Influence Directed Network Model, Music Similarity Model, Genre Analysis Model, Influence Authenticity and Music Revolution Model. The first two models are used to build and refine our network, the latter two models are used to analyze the network. Finally, the model developed is applied to practice. In this part, we will make a summary of our work.

First, a musical influence metric was created to construct a directed network of musical influence. Second, we examined the revolutions and development of musical genres, modeled the similarity of genres and concluded that musicians are more influential within their genre and more similar to other musicians belonging to the same genre. To explore the correlation between genres, hierarchical cluster analysis and time series analysis of genres were used. In terms of the data’s authenticity, it is found that similarity data can suggest that the identified influencers do influence the respective artists. Finally, to find the revolutionaries, Network Analysis, Semantic Analysis, and Random Forest Model are employed.

For the future work, more research is needed in the field of music. Currently, people are becoming more aware of music copyright, yet music cultural appropriation is still very serious. So, can we give deeper features to describe the similarity between music more precisely? In addition, our study found that the popularity of music is positively correlated with the change of time, so does music potentially influence the cognitive structure of human brain and generate new music to drive the development of music?


\section*{Acknowledgments}
We would like to acknowledge the assistance of volunteers in putting
together this example manuscript and supplement.

\bibliographystyle{unsrt}
\bibliography{reference}

\end{document}